\documentclass[useAMS,usenatbib]{mn2e}
\usepackage[dvips]{graphicx}
\usepackage{epsfig}
\usepackage{rotating}
\usepackage{natbib}
\usepackage{color}
\usepackage{mathrsfs}
\usepackage[caption=false]{subfig}

\usepackage{amssymb}

\newcommand{\Msun}{\, {\rm M_{\odot}}}

\newcommand{\Mpc}{\, {\rm Mpc}}

\title[The rates of gas accretion onto galaxies]{The impact of feedback and the hot halo on the rates of gas accretion onto galaxies}
\author[C.A.~Correa et al.]
 {Camila A.~Correa$^{1,2}$\thanks{E-mail: correa@strw.leideuniv.nl}, Joop~Schaye$^1$, Freeke van de Voort$^{3,4}$,
 \newauthor  Alan R.~Duffy$^{5,2}$ and J. Stuart B.~Wyithe$^2$\\
 $^1$ Leiden Observatory, Leiden University, P.O. Box 9513, 2300 RA Leiden, The Netherlands\\
 $^2$ School of Physics, University of Melbourne, Parkville, Victoria 3010, Australia\\
 $^3$ Heidelberg Institute for Theoretical Studies, Schloss-Wolfsbrunnenweg 35, 69118, Heidelberg, Germany\\
 $^4$ Astronomy Department, Yale University, PO Box 208101, New Haven, CT 06520-8101, USA\\
 $^5$ Centre for Astrophysics and Supercomputing, Swinburne University of Technology, Melbourne, Victoria 3122, Australia}
\date{\today}

\pagerange{\pageref{firstpage}--\pageref{lastpage}} \pubyear{2013}

\def\LaTeX{L\kern-.36em\raise.3ex\hbox{a}\kern-.15em
    T\kern-.1667em\lower.7ex\hbox{E}\kern-.125emX}

\begin{document}
\maketitle

\begin{abstract}
We investigate the physics that drives the gas accretion rates onto galaxies at the centers of dark matter haloes using the EAGLE suite of hydrodynamical cosmological simulations. We find that at redshifts $z{\le}2$ the accretion rate onto the galaxy increases with halo mass in the halo mass range $10^{10}-10^{11.7}\Msun$, flattens between the halo masses $10^{11.7}-10^{12.7}\Msun$, and increases again for higher-mass haloes. However, the galaxy gas accretion does not flatten at intermediate halo masses when AGN feedback is switched off. To better understand these trends, we develop a physically motivated semi-analytic model of galaxy gas accretion. We show that the flattening is produced by the rate of gas cooling from the hot halo. The ratio of the cooling radius and the virial radius does not decrease continuously with increasing halo mass as generally thought. While it decreases up to ${\sim}10^{13}\Msun$ haloes, it increases for higher halo masses, causing an upturn in the galaxy gas accretion rate. This may indicate that in high-mass haloes AGN feedback is not sufficiently efficient. When there is no AGN feedback, the density of the hot halo is higher, the ratio of the cooling and virial radii does not decrease as much and the cooling rate is higher. Changes in the efficiency of stellar feedback can also increase or decrease the accretion rates onto galaxies. The trends can plausibly be explained by the re-accretion of gas ejected by progenitor galaxies and by the suppression of black hole growth, and hence AGN feedback, by stellar feedback.
%Stellar feedback also affects the accretion rates onto galaxies. Depending on the halo mass, changes in the efficiency of stellar feedback can either increase or decrease the accretion rates onto galaxies. The trends can plausibly be explained by the re-accretion of gas ejected by progenitor galaxies and by the suppression of black hole growth, and hence AGN feedback, by stellar feedback.
\end{abstract} 


\section{Introduction}

Cosmological simulations have not only shown that the evolution of galaxies' gas reservoirs is governed by feedback from stars and black holes (e.g. \citealt{Keres09,Oppenheimer10,Dubois12,Haas13,Crain17}), but also that it is critically linked to the cosmic web and halo gas flows (e.g. \citealt{Dekel09,vandeVoort12}), which are responsible for the galaxy mass growth. Evidence that feedback manifests itself in the form of enriched outflows and energetic winds comes from various observations (e.g. \citealt{Sharp10,Feruglio10,Cicone14,Anderson15,Turner15,Nielsen16}), but gas accretion from the circumgalactic medium (CGM) is difficult to observe and the physical processes that drive its evolution are not well understood.

It has been proposed that outflows in the form of winds join the warm-hot intergalactic medium and may eventually reverse trajectory to re-accrete onto the galaxy. This is generally referred to as wind recycling or a galactic fountain (see e.g. \citealt{Oppenheimer10,Ubler14,Angles17}). On the other hand, it has also been suggested that gas accretion is suppressed by outflows generated by black holes located in the center of the active galactic nuclei (AGN), that quench the diffuse accretion rates onto galaxies (e.g. \citealt{DiMatteo05,Bower06,Croton,Sijacki,Martizzi,Booth,Dubois12,Dubois}). Accretion shocks occur as a result of collisions between gas from the circum-halo medium falling into haloes and the hot halo gas in hydrostatic equilibrium. When accretion shocks occur, the gravitational energy of the infalling gas is converted into thermal energy. The works of \citet{Rees} and \citet{Silk77} proposed that shock-heated gas with long cooling times forms a hot hydrostatic halo atmosphere, pressure supported against gravitational collapse. 

Semi-analytic models of galaxy formation (SAMs) generally assume that gas falls into galaxies either through a rapid or a slow cooling flow, depending on the radial scale below which gas is able to cool (i.e. where the gas has a cooling time shorter than the dynamical time). This radius is typically referred to as the cooling radius ($r_{\rm{cool}}$). When $r_{\rm{cool}}$ becomes smaller than the virial radius ($R_{200}$), a hot hydrostatic atmosphere is formed. Gas accretion onto galaxies then changes from being in the rapid cooling regime to the slow cooling regime (e.g. \citealt{White91}). Hydrodynamical simulations, however, have shown that both quasistatic and inflowing gas components can exist in the halo at the same radius (see e.g. Figs. 1 and 2 from \citealt{Correa17} and references therein). These are usually referred to as hot and cold modes of accretion. Gas not only falls into a galaxy through a cooling flow (hot mode), but also through a cold flow (cold mode). Cold flows tend to be filamentary, clumpy and of higher density than the hot mode gas, and are strongly correlated with the dark matter filaments that feed haloes (e.g. \citealt{Keres05,Ocvirk,Dekel09,vandeVoort11,Faucher,vandeVoort12,Nelson13,Woods14}). These two modes of accretion are able to coexist in massive haloes at high redshift (\citealt{Dekel09,Correa17}) and feed the galaxy at the same time. The co-evolution between the different modes of accretion has not been implemented in SAMs until recently. \citet{Cousin15} included a two phase smooth baryonic accretion, with the hot and cold component built over the smooth dark matter accretion, whereas \citet{Lu15} modelled a circum-halo medium, assumed to be preheated up to a certain entropy level, to reduce the baryonic accretion. 

Once gas crosses the hot halo and cools, feedback from stars and black holes can potentially reheat it and prevent it from falling into the galaxy. Therefore, due to the complexity in the interaction of all the possible mechanisms that modify the manner in which galaxies accrete gas, a physical model of galaxy gas accretion is still missing. In this work, we use the ``Evolution and Assembly of GaLaxies and their Environments'' (EAGLE) simulations to study the modes of gas accretion onto galaxies and their dependence on feedback variations. We derive a physically motivated model of galaxy gas accretion that aims to explain the underlying physics of the way gas cools from the hot halo and accretes onto galaxies. In \citet[hereafter Paper I]{Correa17}, we derived a new criterion to determine when the hot halo forms, based on the fraction of gas accreting onto haloes that shock-heats to the halo virial temperature (i.e. hot accretion) and on the hot halo gas mass. We calculated a heating rate produced by accretion shocks and compared it to the gas cooling rate. We found that haloes with masses above the critical mass threshold of $10^{11.7}\Msun$ are able to develop a hot stable hydrostatic atmosphere, in agreement with previous work (e.g. \citealt{Birnboim,Dekel}). In Paper I we showed that feedback affects the mass-scale of hot halo formation and impacts the distribution of gas in the halo. In this work, we make use of the analytic heating and cooling rates from Paper I and assume that there are two modes of galaxy gas accretion, hot and cold.  

This work is organized as follows. In Section~\ref{Simulations_sec} (as well as in Appendix~\ref{Simulations_app}) we describe the numerical simulations used and the methods we employ to measure gas accretion rates. In Section~\ref{gas_accretion_sec} we analyze the total gas accretion rates onto galaxies from the EAGLE simulations as a function of halo mass and redshift, as well as the hot and cold modes of accretion (Section~\ref{hot_cold_modes_of_accretion}), and the impact of stellar and AGN feedback (Sections~\ref{stellar_fb} and \ref{AGN_fb}, respectively). In Section~\ref{AGN_hot_halo_sec} we derive a physically motivated model of gas accretion onto galaxies, we show that the model reproduces the gas accretion rates from the EAGLE simulations, and we analyze the role of AGN in the rates of gas cooling from the hot halo. Finally, we summarize and conclude in Section~\ref{Conclusion_sec}.

\begin{table*}
\centering  %r@{.}l
\caption{List of simulations. From left-to-right the columns show: simulation name; comoving box size; number of dark matter particles (there are equally many baryonic particles) and brief description.} 
\label{Table_sims2}
\begin{tabular}{lrrl}
\hline
  Simulation name & L ($\rm{cMpc}$) & N & Description \\  \hline\hline
  Ref & 100 & $1504^{3}$ & ref. stellar \& AGN feedback\\
  No AGN FB & 50 & $752^{3}$ & ref. stellar \& no AGN feedback\\
  More Explosive AGN FB & 50 & $752^{3}$ & ref. stellar \& more explosive and intermittent AGN feedback\\ 
  & & & (but same energy injected per unit mass accreted by the BH as Ref)\\
  More Energetic Stellar FB & 25 & $376^{3}$ & twice as much energy injected per unit stellar mass with respect to\\
  & & & Ref \& ref. AGN feedback\\
  Less Energetic Stellar FB & 25 & $376^{3}$ & half as much energy injected per unit stellar mass with respect to\\
  & & & Ref \& ref. AGN feedback\\
  No Stellar FB & 25 & $376^{3}$ & no stellar feedback \& ref. AGN feedback\\
  No Stellar/AGN FB & 25 & $376^{3}$ & no stellar feedback \& no AGN feedback\\\hline
\end{tabular}
\end{table*}

\section{Simulations}\label{Simulations_sec}

The EAGLE simulation suite (\citealt{Schaye14,Crain15}) was run using a modified version of GADGET-3 (\citealt{Springelb}), an $N$-Body Tree-PM smoothed particle hydrodynamics (SPH) code, but with a new formulation of SPH, new time stepping and new subgrid physics. The simulations assume a $\Lambda$CDM cosmology with the parameters given by {\it{Planck-1}} data (\citealt{Planck}), $\Omega_{\rm{m}}=1-\Omega_{\Lambda}=0.307$, $\Omega_{\rm{b}}=0.04825$, $h=0.6777$, $\sigma_{8}=0.8288$, $n_{s}=0.9611$, and are run from redshift $z=127$ to $z=0$. Throughout this work we use simulations with different box sizes (ranging from 25 to 100 comoving $\Mpc$) and particle numbers (ranging from $2\times 376^{3}$ to $2\times 1504^{3}$), but the same resolution. For clarity, the simulation names contain strings of the form $L$xxx$N$yyyy, where xxx is the simulation box size in comoving $\Mpc$ and yyyy is the cube root of the number of particles per species (where the number of baryonic particles is equal to the number of dark matter particles). 

For our analysis we mainly use the L100N1504 reference model (hereafter Ref), that contains an initial baryonic particle mass of $1.81\times 10^{6}\Msun$, a dark matter particle mass of $9.70\times 10^{6}\Msun$, a comoving (Plummer equivalent) gravitational softening of 2.66 comoving kpc and a maximum physical softening of 0.7 proper kpc. Although the Ref model was calibrated without considering gas properties, it is able to reproduce the cosmic HI column density distribution and circumgalactic covering fraction profiles (\citealt{Rahmati15,Rahmati16}) and the neutral gas mass and profiles (\citealt{Bahe16}). Note, however, that the HI masses of present-day dwarf galaxies are underestimated (\citealt{Crain17}).

In order, to investigate the impact of feedback on the rates of galaxy gas accretion, we additionally use simulations with varying AGN and stellar feedback prescriptions. In the stellar feedback case, we use simulations with `Less/More Energetic Stellar FB', which means that the energy injected per unit mass of stars formed is half/twice the amount used in Ref. In the AGN case, we use `No AGN FB' and `More Explosive AGN FB', meaning that AGN feedback is switched off or is more explosive and intermittent than the Ref model, respectively (but note that in the latter case the energy injected per unit mass accreted by the BH does not change with respect to Ref). See Table~\ref{Table_sims2} for reference. In Appendix A (also in Paper I) we include a brief description of the modeling of the EAGLE simulations, but see \citet{Schaye14} and \citet{Crain15} for more details. The EAGLE simulations are publicly available; for details see \citet{McAlpine16}.

Dark matter haloes in EAGLE (and the self-bound substructures within them associated with galaxies) are identified using the Friends-of-Friends (FoF) and SUBFIND algorithms (\citealt{Springel,Dolag}). The virial masses and radii are determined using a spherical overdensity routine within SUBFIND, centered on the minimum gravitational potential of the main subhalo from the FoF group. Halo masses ($M_{200}$) are defined as the total mass within a radius, $R_{200}$, within which the mean density is 200 times the critical density. In each FoF halo, the `central' galaxy is the galaxy closest to the center (minimum of the potential), which is usually the most massive. The remaining galaxies within the FoF halo are its satellites. Throughout this work we focus on the gas accretion rates onto central galaxies. The gas accretion rates onto satellite galaxies differ from the rates onto centrals because they are strongly suppressed by environmental effects, such as ram pressure or tidal stripping. For details of these effects on gas accretion rates onto satellite galaxies from EAGLE see \citet{vandeVoort17}.

\subsection{Methodology}\label{Methodology}

In this section we describe the methodology we employ to calculate the gas accretion rates onto galaxies at the center of dark matter haloes. We begin by building merger trees across the simulation snapshots\footnote{The simulation data is saved in 9 discrete output redshifts between redshift 0 to 1, in 8 output redshifts between redshift 1 and 3, and in 8 output redshifts between redshift 3 and 8.}. At each output redshift we select haloes that contain more than 1000 dark matter particles (which corresponds to a minimum halo mass of $M_{200}=10^{9.8}\Msun$ in the Ref-L100N1504 simulation). This particle number-cut is based on the convergence analysis presented in Appendix \ref{Numerical_convergence}, where we find that in smaller haloes the accretion onto galaxies does not converge, indicating that the inner galaxies are not well resolved (see Appendix \ref{Numerical_convergence} for a discussion). We refer to these haloes as `descendants', and link each descendant with a unique `progenitor' at the previous output redshift. This is nontrivial due to halo fragmentation: subhalos of a progenitor halo may have descendants that reside in more than one halo. The fragmentation can be spurious or due to a physical unbinding event. To correct for this, we link the descendant to the progenitor that contains the majority of the descendant's 25 most bound dark matter particles (see \citealt{Correa15b} for an analysis of halo mass history convergence using the mentioned criteria to connect haloes between snapshots). To calculate the gas accretion onto haloes, we perform a particle ID matching between particles within linked haloes from consecutive snapshots. Particles that are new to the system, and are within the virial radius, are labeled as accreted particles in the redshift range $z_{i} < z < z_{j}$. 

Different methods have been employed to determine gas accretion onto galaxies. For example, \citet{Faucher} measured accretion rates through shells of prescribed radii. In order to differentiate outflows from inflowing material, they added the particles within a given shell and defined the net accretion as $\dot{M}\propto \sum_{p} M_{p}{\bf{v}}_{p}/\Delta r_{p}$ (with $M_{p}$ the particle mass, $\bf{v}_{p}$ the velocity vector and $\Delta r_{p}$ the thickness of shell). They classified the net accretion rates as inward or outward according to the direction of the velocity vector. A different approach was used by \citet{vandeVoort17} who, in order to separate the galaxy from the halo, used a radial cut of 30 pkpc and a cut in the hydrogen number density ($n_{\rm{H}}> 0.1\hspace{1mm}\rm{cm}^{-3}$) to define the interstellar medium (ISM) (see also \citealt{vandeVoort11}). They considered particles that are star-forming and part of the ISM at $z_{i}$, but which were gaseous and not part of the ISM at $z_{j}$ to have been accreted onto a galaxy at $z_{i} < z < z_{j}$. Similarly, \citet{Nelson13} made use of a density-temperature ($\rho-T$) cut criterion ($\log_{10}(T/{\rm{K}})-0.25\log_{10}(\rho/\rho_{{\rm{crit}},z=0}) < 4.11$, with $\rho_{{\rm{crit}},z=0}$ the critical density today), along with a radial cut (${<0.15\times R_{200}}$). They considered a gas element to have accreted onto a galaxy if it belonged to that galaxy at $z_{j}$, and either crossed the phase space cut in $\rho-T$ or the radial cut during $z_{i} < z < z_{j}$.

In this work we closely follow the method of \citet{vandeVoort17} and define the ISM to consist of all particles within a sphere of radius $0.15\times R_{200}$ (centered on the minimum of the gravitational potential) that are either star-forming (i.e. have $n_{\rm{H}}>n^{*}_{\rm{H}}=0.1{\rm{cm}}^{-3}(Z/0.002)^{-0.64}$, with $Z$ metallicity, and $T<10^{0.5}T_{\rm{EoS}}$, with $T_{\rm{EoS}}\propto \rho^{1/3}$ a temperature floor that corresponds to the equation of state and is normalised to $T_{\rm{eos}}=8\times 10^{3}$K at $n_{\rm{H}}=10^{-1}$cm$^{-3}$) or part of the atomic ISM (i.e. have $n_{\rm{H}}>0.1\hspace{1mm}{\rm{cm}}^{-3}$ and $T<10^{5}\hspace{1mm}\rm{K}$). We next calculate three rates of gas accretion onto $0.15\times R_{200}$: gas accretion of all gas particles that cross the $0.15\times R_{200}$ radius between $z_{i} < z < z_{j}$ (hereafter $\dot{M}_{0.15R_{200}}$), gas accretion of only star-forming particles ($\dot{M}_{\rm{SFR}>0}$) and gas accretion of only gas particles that form part of the ISM ($\dot{M}_{\rm{ISM}}$). In all the rates we are including star particles that were gas particles in the previous output but turned into stars during the time step between the snapshots. 

Note that $\dot{M}_{\rm{SFR}>0}$ and $\dot{M}_{\rm{ISM}}$ refer to gas accretion onto the star forming and ISM components within $0.15\times R_{200}$, and that $\dot{M}_{\rm{ISM}}$ includes $\dot{M}_{\rm{SFR}>0}$. Also note that the accreting gas is possibly, but not necessarily, star forming prior to accretion. We further define $\dot{M}_{\rm{0.15R_{200}}}$ as the rate of gas accretion onto galaxies (hereafter $\dot{M}_{\rm{gas,galaxy}}=\dot{M}_{\rm{0.15R_{200}}}$), because we assume that the `galaxy' extends beyond the ISM. 

To summarize, $\dot{M}_{\rm{gas,galaxy}}$ refers to the accretion rates of gas that crosses the radial boundary $0.15\times R_{200}$ during $z_{i} < z < z_{j}$, $\dot{M}_{\rm{ISM}}$ considers gas that crosses the radial boundary and the phase space cut $n_{\rm{H}}-T$ or is star-forming during $z_{i} < z < z_{j}$, and $\dot{M}_{\rm{SFR}>0}$ considers gas that crosses the radial boundary during $z_{i} < z < z_{j}$ and is star-forming at $z_{i}$. Note that in each case we calculate the accretion rates by adding the mass of all the accreted particles (following the condition of accretion) onto individual galaxies and dividing by the time interval (that is $\sim 1.34$ Gyr at $z=0$ and $\sim 0.33$ Gyr at $z=2$). The final rates given are the median accretion rates onto galaxies in bins of halo mass. Note that we calculate the `net' rate of gas accretion between two consecutive snapshots, i.e. we do not separate gas that is accreted for the first time or re-accreted. 

Gas particles that fall into galaxies can be classified as gas introduced to the system by a merger event or through smooth accretion. Mergers have been defined as the accretion of particles that belonged to any resolved subhalo at the previous snapshot (e.g. \citealt{Keres09,Nelson13}) or to subhalos with masses above 1/10th of main progenitor mass (e.g. \citealt{vandeVoort11}). It has been found that the specific gas accretion rate onto haloes through mergers is much lower than the specific smooth accretion rate, except for high-mass haloes ($M_{200}\gtrsim10^{14}\Msun$) at $z=0$ (\citealt{vandeVoort11}). In the case of galaxies at $z=2$, gas from all resolved merger events can contribute as much as $60\%$ to the total galaxy accretion rate at all halo masses (\citealt{Nelson13}). At lower redshifts the contribution decreases and the gas supply from resolved mergers appears to be only important at the high-mass end (\citealt{Keres09}). In this work however, we do not focus on the origin of the gas that falls into galaxies. Instead we investigate the physics that prevents gas from cooling.

%The origin of the gas particles that fall into galaxies is beyond the scope of this work. Previous works have analysed the contribution of mergers and smooth accretion to the total galaxy accretion rates (Ocvirk et al. 2008; Keres et al. 2009; Brooks et al. 2009; van de Voort et al. 2011; Nelson et al. 2013). Mergers are generally defined as accreted particles that reside in subhalos above some minimum mass at the previous snapshot. 

%For instance, van de Voort et al. (2011) compared the two and showed that the specific gas accretion rate through mergers is much lower than the specific smooth accretion rate, except for high-mass haloes ($M_{200}=10^{14}\Msun$) at $z=0$. Although the exact approach in separating out the merger contribution differs, it has been found that at redshift $z=2$ gas from merger events can contribute to as much as $60\%$ to the total galaxy accretion rate at all halo masses (Nelson et al. 2013). Al lower redshifts the contribution decreases and the gas supply from resolved mergers appears to be only important at the high-mass end (Keres et al. 2009).

\begin{figure} 
\centering
  \includegraphics[angle=0,width=0.46\textwidth]{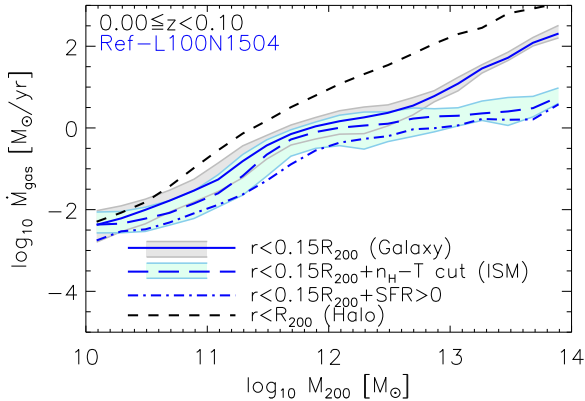}\\
  \vspace{-0.1cm}
  \caption{Accretion rate of gas onto the central galaxies of dark matter haloes as a function of halo mass in the redshift range $0\leq z<0.1$. The solid line corresponds to the gas accretion rate calculated by counting all gas particles that crossed $0.15\times R_{200}$ during consecutive snapshots, whereas the dashed and dot-dashed lines correspond to the gas accretion rates calculated by counting gas particles that crossed $0.15\times R_{200}$ radius and the phase-space cut $n_{\rm{H}}-T$, and that crossed the $0.15\times R_{200}$ radius and are star-forming, respectively. The grey and cyan shaded regions enclosing the median values of $\dot{M}_{\rm{gas}}$ correspond to the $1\sigma$ scatter (16-84th percentiles) onto the $0.15\times R_{200}$ region and ISM, respectively. The $1\sigma$ scatter of $\dot{M}_{\rm{SFR}>0}$, not included in the figure, is (on average) 0.3 dex similar to that for $\dot{M}_{\rm{ISM}}$. For comparison, the black dashed line shows the rate of gas accretion onto haloes (i.e. crossing $R_{200}$). The figure shows that massive galaxies accrete warm diffuse gas that is not accreted onto the ISM.}
\label{growthrate_plot}
\end{figure}

Fig.~\ref{growthrate_plot} shows $\dot{M}_{\rm{gas,galaxy}}$ (blue solid line), $\dot{M}_{\rm{ISM}}$ (blue long-dashed line) and $\dot{M}_{\rm{SFR}>0}$ (blue dot-dashed line), as a function of halo mass, taken from the Ref-L100N1504 simulation. For comparison, the figure also shows the gas accretion rate onto haloes (i.e. crossing $R_{200}$, black dashed line). In the halo mass range $10^{10}-10^{12}\Msun$, $\dot{M}_{\rm{gas,galaxy}}$ and $\dot{M}_{\rm{ISM}}$ increase with halo mass at approximately the same rate, with $\dot{M}_{\rm{ISM}}$ having a 0.3 dex (on average) lower normalization than $\dot{M}_{\rm{gas,galaxy}}$. For halo masses greater than $10^{12}\Msun$, $\dot{M}_{\rm{ISM}}$ remains roughly constant. While $\dot{M}_{\rm{gas,galaxy}}$ also flattens in $10^{12}\Msun$ haloes, it increases with halo mass for $\gtrsim 10^{13}\Msun$ haloes. $\dot{M}_{\rm{SFR}>0}$ behaves similarly to $\dot{M}_{\rm{ISM}}$ though it is (on average) a factor of 2 lower. In the figure, the grey and cyan shaded regions enclosing the median values of $\dot{M}_{\rm{gas,galaxy}}$ and $\dot{M}_{\rm{ISM}}$ correspond to the $1\sigma$ scatter (16-84th percentiles). 

From Fig.~\ref{growthrate_plot} it can be seen that in the halo mass range $10^{10}-10^{12}\Msun$, roughly $50\%$ of the gas crossing $0.15R_{200}$ joins the ISM. At higher halo masses, the fraction of gas crossing $0.15R_{200}$ that joins the ISM decreases and most of the gas that reaches the inner halo has either a low density ($n_{H}<0.1$ cm$^{-3}$) or a high temperature ($>10^{5}$ K). We then conclude that massive galaxies accrete warm diffuse gas that does not fall into the ISM.

In the following sections we investigate the way gas crosses the CGM and the origin of $\dot{M}_{\rm{gas,galaxy}}(M_{200})$ further by first disentangling the impact of stellar and AGN feedback (Sections \ref{stellar_fb} and \ref{AGN_fb}, respectively). In addition, we calculate the fraction of hot/cold modes of gas accretion onto galaxies in Section~\ref{hot_cold_modes_of_accretion}. To estimate these modes of gas accretion, previous studies followed the temperature history of the accreted gas (see e.g. \citealt{Faucher,vandeVoort11}), however, in Paper I we concluded that selecting gas particles according to their temperature just after accretion, $T_{\rm{post-shock}}$, is a better method to determine hot/cold accretion. This is because it excludes gas particles that go through a shock but immediately cool afterwards, or that did not pass through an accretion shock but instead were heated in the past by stellar feedback and have since cooled. Then, throughout this work, we define the fraction of gas particles accreted hot, $f_{\rm{acc,hot}}$, as the fraction of particles that after being accreted have temperatures larger than $T_{\rm{post-shock}} = 10^{5.5}\hspace{1mm}\rm{K}$. Note that for low-mass haloes the results are sensitive to this specific temperature threshold (e.g. \citealt{vandeVoort11}). For an analysis of how the hot mode of accretion depends on the temperature threshold, as well as on other criteria (e.g. $T_{\rm{vir}}$, cooling times), see Sections 2.2 and 4 of Paper I.

\begin{figure*} 
  \centering
  \vspace{-0.3cm}
  \subfloat{\includegraphics[angle=0,width=0.405\textwidth]{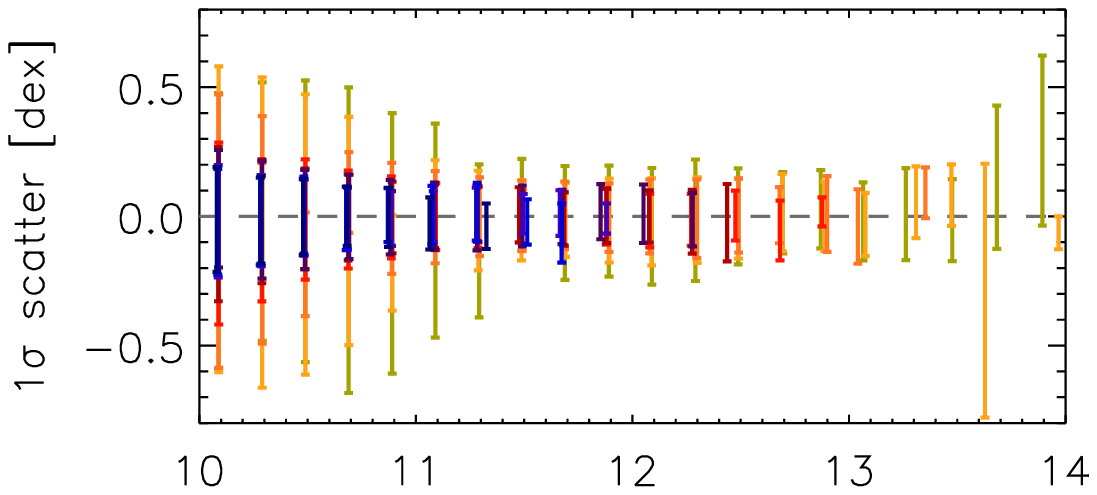}}
  \hspace{-0.1cm}
  \subfloat{\includegraphics[angle=0,width=0.405\textwidth]{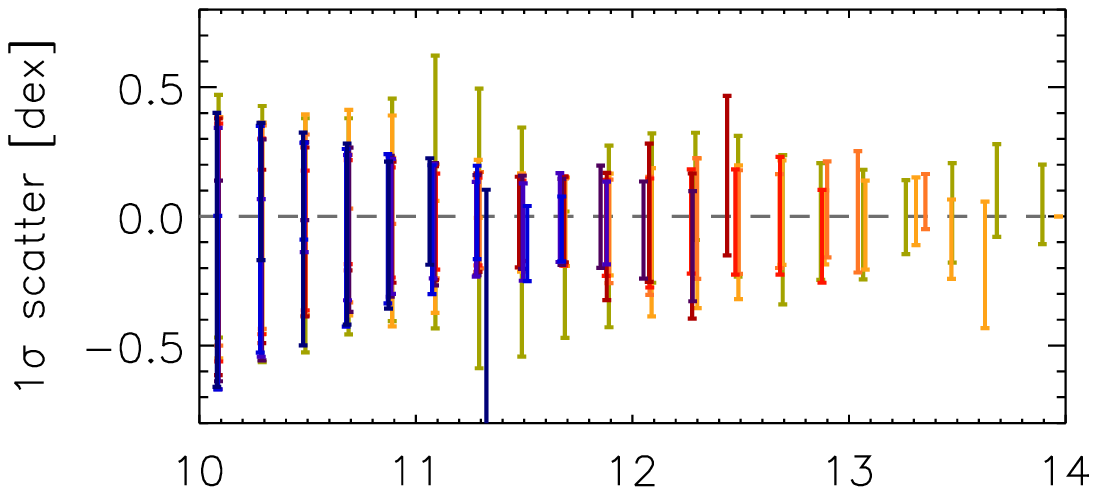}}\\
  \vspace{-0.4cm}
  {\hspace{0.15cm}
  \subfloat{\includegraphics[angle=0,width=0.39\textwidth]{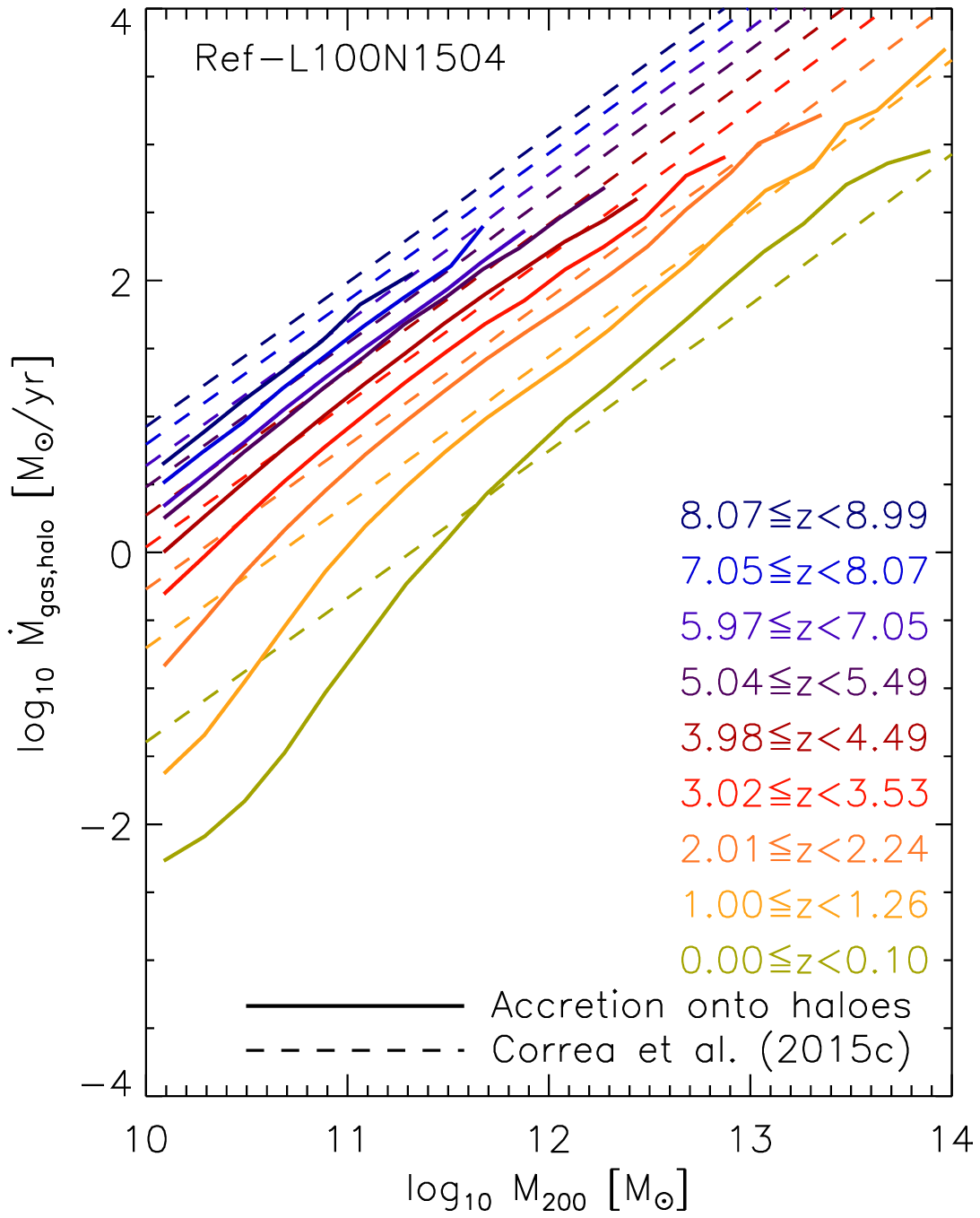}}}
  {\hspace{0.1cm}
  \subfloat{\includegraphics[angle=0,width=0.39\textwidth]{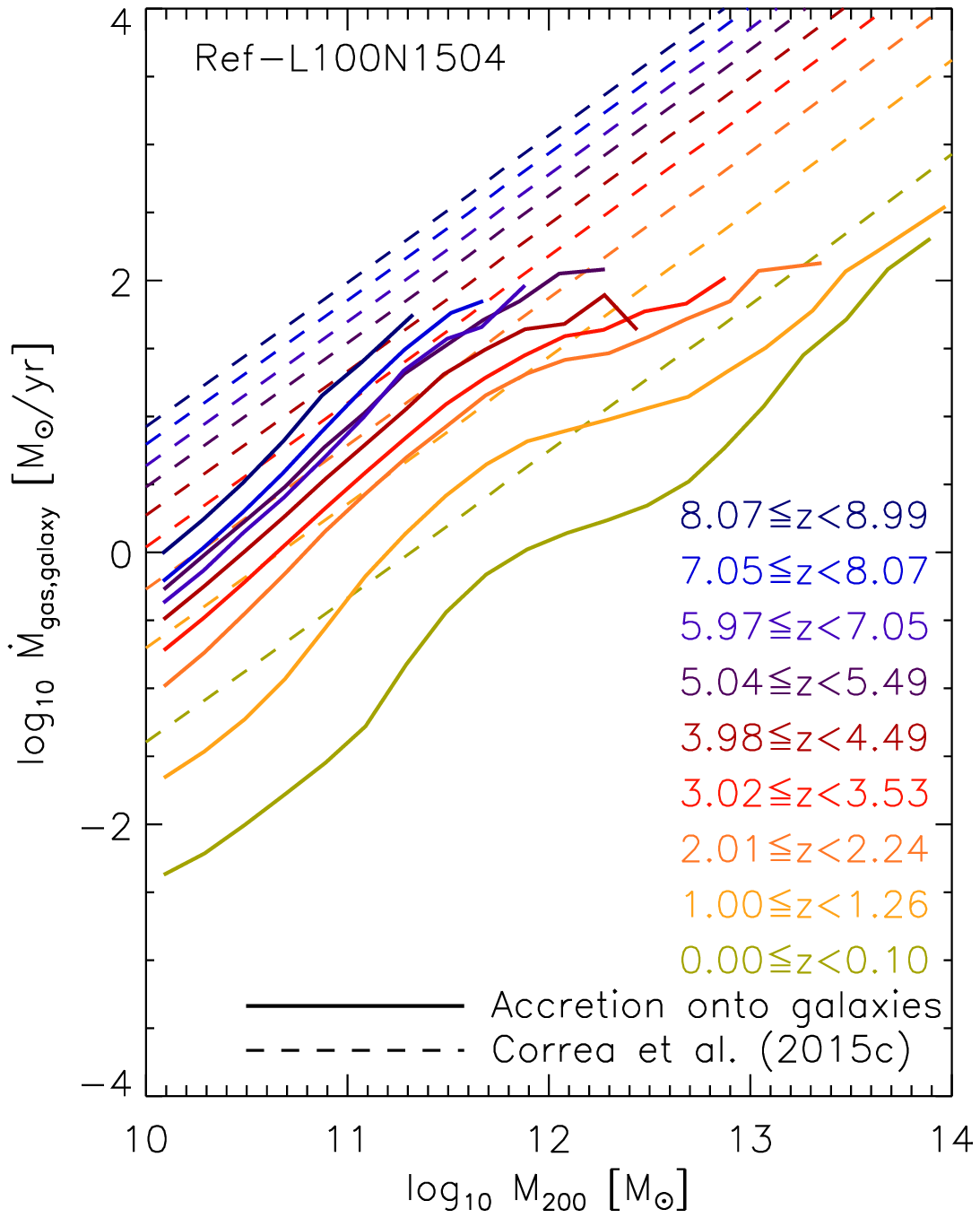}}}\\
  \vspace{-0.15cm}
  \caption{Accretion rates onto haloes (bottom-left panel) and galaxies (bottom-right panel) as a function of halo mass for different redshifts. The curves are colored according to the redshift intervals indicated in the legends. The solid curves correspond to the median accretion rates taken from the Ref-L100N1504 simulation, whereas the dashed curves correspond to the analytic model of \citet{Correa15c} for dark matter halo accretion rates times $f_{\rm{b}}=\Omega_{\rm{b}}/\Omega_{\rm{m}}$. The $1\sigma$ scatter (16-84th percentiles) of the median accretion rates onto haloes and galaxies are shown in the top-left and top-right panel, respectively. The figure shows that at $z\leq 2$ in the halo mass range $10^{10}-10^{12}\Msun$ the gas accretion rates onto the galaxy increases with halo mass but flattens at ${\sim}10^{12}\Msun$.}
\label{accretion_plot}
\end{figure*}

\section{Gas accretion rates}\label{gas_accretion_sec}

In this section we calculate the rates of gas accretion onto haloes and galaxies from the EAGLE simulations. For a detailed analysis of the numerical convergence of our results, see Appendix~\ref{Numerical_convergence}. 

%We find excellent agreement between the accretion rates onto haloes and galaxies from simulations with same resolution and different box size, but we do not achieve `strong convergence', which means agreement between simulations with different numerical resolution and same parameters of the subgrid models. In the latter case we find that in $10^{11}\Msun$ haloes, $\dot{M}_{\rm{gas,galaxy}}$ increases by up to a factor of 10 if the mass resolution is increased by a factor of 64, whereas $\dot{M}_{\rm{gas,halo}}$ increases by up to a factor of 2.5. This is not discouraging because we achieve `weak convergence', meaning that we find excellent agreement in the accretion rates between the recalibrated model `Recal-L025N0752' and the reference models Ref-L100N1504/Ref-L025N0376. As explained in \citet{Schaye14}, although strong convergence is a necessary condition for predictive power, only weak convergence is required for observables that depend strongly and directly on the efficiency of subgrid feedback. For a detailed analysis of numerical convergence see Appendix~\ref{Numerical_convergence}. 

\subsection{Accretion rates onto galaxies and haloes}\label{accretion_rates_halos_galaxies}

Fig.~\ref{accretion_plot} shows the total gas accretion rate onto haloes (bottom-left panel) and onto galaxies (bottom-right panel) over many redshift intervals. The solid lines correspond to the median gas accretion rates from the Ref-L100N1504 simulation. The figure shows the median accretion rates of haloes separated in logarithmic mass bins of 0.2 dex. The dashed lines correspond to the analytic accretion rates from \citet{Correa15a,Correa15c} multiplied by the universal baryon fraction ($f_{\rm{b}}=0.157$). In the figure, all the curves are coloured according to the redshift interval, as indicated in the legends. The top panels show the $1\sigma$ scatter of $\dot{M}_{\rm{gas,halo}}$ (top-left) and $\dot{M}_{\rm{gas,galaxy}}$ (top-right) for each mass bin.

%In this case, we are not using the universal baryon fraction ($f^{\rm{univ}}_{b}=\Omega_{\rm{b}}/\Omega_{\rm{m}}=0.157$). Instead, we find that in order to obtain a good agreement between the curves at high redshift, $f_{\rm{b}}$ must be $f_{\rm{b}}=0.1$. This is in agreement with \citet{Schaller15}, who calculated the mass fractions of baryons within $R_{200}$ and found that the baryon fractions are much lower than the universal value for all halos smaller than $10^{14}\Msun$, i.e. $f_{\rm{b}}\approx 0.05$ for $10^{12}\Msun$ halos and $f_{\rm{b}}\approx 0.1$ for $10^{13}\Msun$ halos (see also \citealt{Qin17}). 

We find that the $z=0$ accretion rate onto haloes deviates from the analytic prediction of \citet{Correa15c}. The prediction is 0.8 dex too high for $10^{10}\Msun$ haloes, agrees for $10^{11.5}\Msun$ haloes, and is 0.3 dex too low for $10^{13}\Msun$ haloes. In the redshift range $z=1-2$ the prediction is also $\sim 0.8$ dex too high for $10^{10}\Msun$ haloes, but a better agreement is reached for haloes more massive than $10^{12}\Msun$. At higher redshifts the disagreement between the analytic prediction and the simulation output increases, with the prediction being up to 0.5 dex too high for all halo masses at $z=8$.

A possible explanation for the difference in the halo accretion rates between the Ref model (solid lines in left panel of Fig.~\ref{accretion_plot}) and the analytic prediction of \citet{Correa15c} (dashed lines) is that the analytic model does not consider the impact of baryon physics (such as gas pressure, cooling, reionization and stellar and AGN feedback), that reduces the halo mass by a factor of 0.7 for $10^{10}\Msun$ haloes at $z=0$ (\citealt{Schaller15}) and $\sim 0.6$ for all haloes at $z>6$ (\citealt{Qin17}). However, in low-mass haloes the disagreement between the analytic model and the simulation output is expected, because at these masses the extragalactic UV/X-ray background radiation heats the surrounding gas and prevents it from falling into the halo (e.g. \citealt{Sawala13,Benitez17}).

We look for the best-fit expression that reproduces the halo gas accretion rates in the presence of feedback from the Ref model. We find it to be

\begin{eqnarray}\label{halo_accr1}
\log_{10}\dot{M}_{\rm{gas,halo}} &=& a_{z\le 4}(z)+b_{z\le 4}(z)x+c_{z\le 4}(z)x^{2},\\\label{halo_accr2}
a_{z\le 4}(z)&=&0.830+0.553z-0.0523z^2,\\\label{halo_accr3}
b_{z\le 4}(z)&=&1.436-0.149z+0.007z^2,\\\label{halo_accr4}
c_{z\le 4}(z)&=&-0.134+0.099z-0.023z^2,\\\label{halo_accr5}
x &=& \log_{10}(M_{200}/10^{12}\Msun),
\end{eqnarray}

\noindent if $z \le 4$ and

\begin{eqnarray}\label{halo_accr6}
\log_{10}\dot{M}_{\rm{gas,halo}}&=& a_{z>4}(z)+b_{z>4}(z)x,\\\label{halo_accr7}
a_{z> 4}(z)&=&3.287-0.401z+0.045z^2,\\\label{halo_accr8}
b_{z> 4}(z)&=&1.016+0.003z+0.002z^2.
\end{eqnarray}

\noindent if $z>4$. We show a comparison between the best-fit expression and simulation output in Appendix~\ref{Comparison}.

The bottom-right panel of Fig.~\ref{accretion_plot} shows that the gas accretion rate onto galaxies is much lower than that onto haloes. The dependence on mass is also quite different. Although the galaxy accretion rate initially increases with halo mass, it flattens in the halo mass range $10^{11.7}-10^{12.7}\Msun$, particularly at $z\le 2$. We believe that the flattening is produced by the presence of the hot halo atmosphere, which forms in $10^{11.7}\Msun$ haloes (Paper I). In Section~\ref{model_section} we investigate this further by deriving an analytic model that includes the heating and cooling rates of gas from the hot halo.

The galaxy accretion rates calculated in this work differ from those found in other simulations (\citealt{Keres05,Ocvirk,Faucher,vandeVoort11,Nelson13}). To mention a few estimates, \citet{vandeVoort11} obtained accretion rates onto the ISM of $\sim 1$, $10$ and $15\Msun$/yr in $10^{12}$, $10^{12.5}$ and $10^{13}\Msun$ haloes, respectively, at $z=0$. \citet{Faucher} found rates of cold accretion (gas particles with hydrogen number density lower than $0.13$ $\rm{cm}^{-3}$ and past maximum temperature lower than $2.5\times 10^{5}$ K) onto $0.2R_{200}$ of 0.3 and 1$\Msun$/yr in $10^{12}$ and $10^{13}\Msun$ haloes at $z=0$, respectively. Similarly, \citet{Nelson13} obtained total rates of galaxy smooth gas accretion of 1 and 5$\Msun$/yr in $10^{11}$ and $10^{12}\Msun$ haloes, respectively, at $z=0$. As discussed in Section 2.1, these works differ on the method employed to calculate the rates of gas accretion, therefore we do not expect good agreement, but we find that our results closely follow those of \citet{vandeVoort11}. 

Note that the rates of galaxy and halo gas accretion in this work are calculated counting new gas particles within $0.15\times R_{200}$ and $R_{200}$, respectively. It is then possible that halo pseudo-evolution (the growth in halo mass due to the redshift evolution of the reference density, e.g. \citealt{Diemer}) affects the rates of accretion. This occurs if gas particles not inflowing relative to the physical density profile of the halo appear to accrete when $R_{200}$ increases due to pseudo-evolution from one snapshot to the next. We next analyse this possibility. In this work, we calculate the gas accretion rates within time intervals of up to $\approx 1.3$Gyr (e.g. between redshifts 0 and 0.1). We test this by calculating the gas accretion rates onto a fixed proper radius given by $r_{\rm{gal}}=0.15R_{200}(z=0)$. We obtain that when the radius is kept fixed, the accretion rate decreases (on average) $9\%$ for $10^{10}-10^{13}\Msun$ haloes, to a maximum of $13\%$ for $10^{11.6}\Msun$ haloes. We conclude that the change in radius due to pseudo-evolution is not large enough to affect our results significantly.

\begin{figure} 
 \centering
  \subfloat{\includegraphics[angle=0,width=0.49\textwidth]{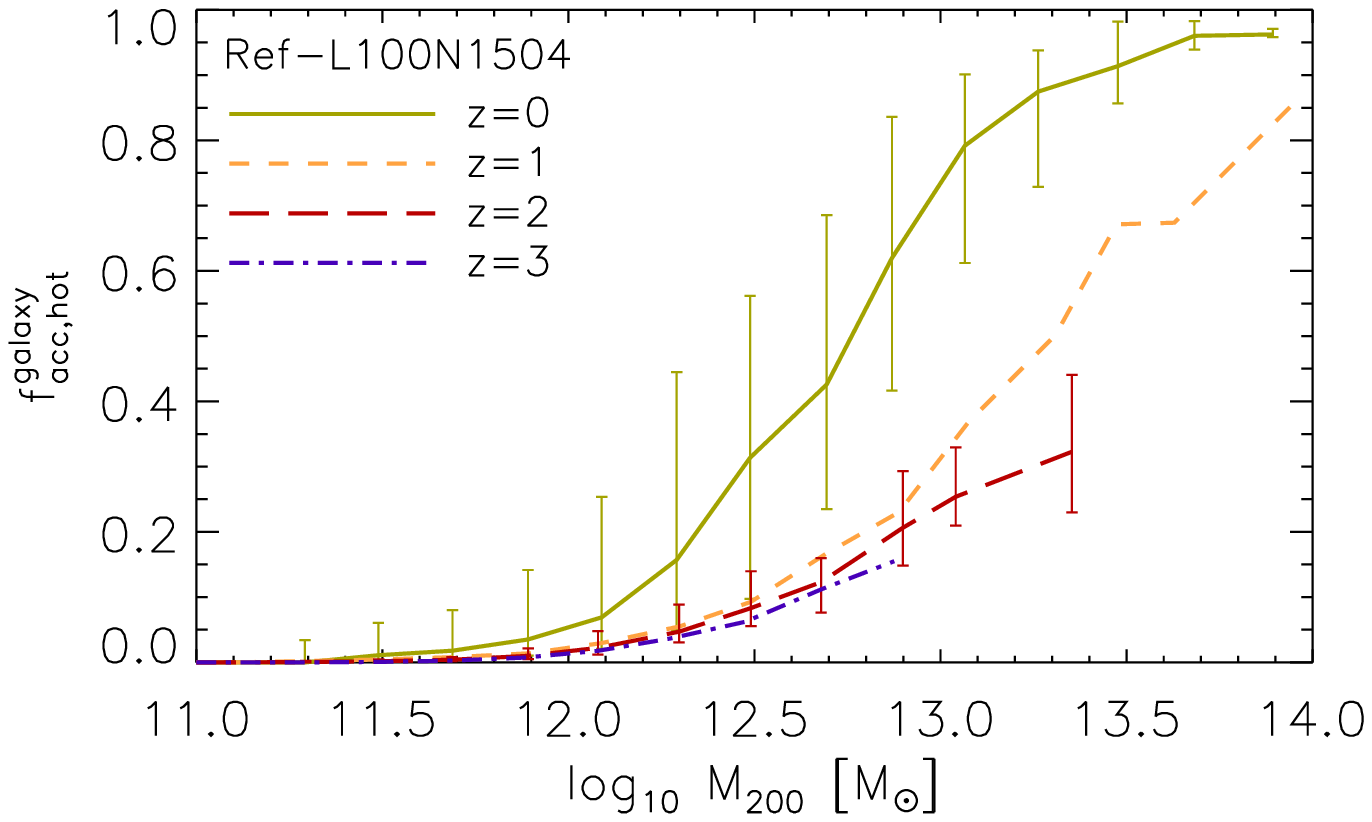}}\\
  \vspace{-0.66cm}
  \subfloat{\includegraphics[angle=0,width=0.49\textwidth]{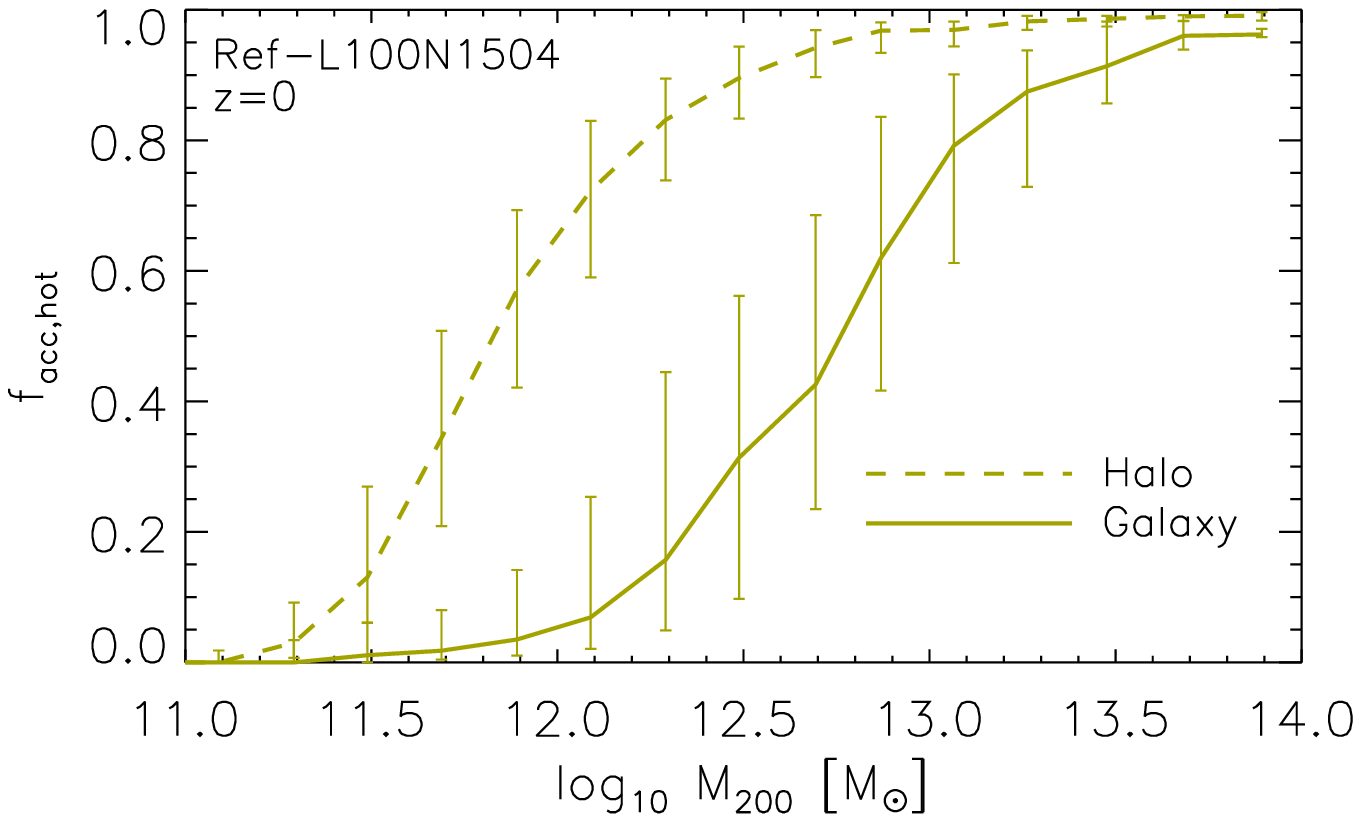}}
    \vspace{-0.1cm}
  \caption{Top panel: median fraction of gas accreted in the hot mode onto galaxies as a function of halo mass in the redshift ranges $0\leq z<0.1$ (green solid line), $1\leq z<1.26$ (orange dashed line), $2\leq z<2.24$ (red long-dashed line) and $3\leq z<3.53$ (purple dot-dashed line). Bottom panel: Fraction of gas accreted in the hot mode onto galaxies (solid line) and onto haloes (dashed line) as a function of halo mass in the redshift range $0\leq z<0.1$. The error bars in the figure show the $1\sigma$ scatter. We find that while hot accretion onto haloes dominates in haloes more massive than $10^{11.8}\Msun$, hot accretion onto galaxies only dominates for haloes more massive than $>10^{12.8}\Msun$ at $z=0$.}
\label{accretion_hotcold_plot}
\end{figure}

\subsection{Hot and cold modes of accretion}\label{hot_cold_modes_of_accretion}

%When gas crosses the virial radius it can either experience a shock and be heated to the halo virial temperature or infall into the halo unperturbed. 
In Paper I we calculated the fractions of hot and cold modes of gas accretion onto haloes using the EAGLE simulations, and found that the hot fraction increases smoothly with halo mass (from 0.1 for $10^{11.5}\Msun$ haloes to 0.7 for $10^{12}\Msun$ haloes at $z=0$), and decreases with increasing redshift (from 0.5 in $10^{12}\Msun$ haloes at $z=1$ to 0.2 at $z=2$). In this section we extend this analysis by focusing on the modes of gas accretion onto galaxies. We select gas particles using the radial cut ($0.15\times R_{200}$) and separate the hot and cold modes by applying a temperature cut of $10^{5.5}$ K (see Section~\ref{Methodology} for a description on how we calculate hot/cold gas accretion). Throughout this work we label the fractions of hot/cold modes of gas accretion onto galaxies as $f^{\rm{galaxy}}_{\rm{acc,hot/cold}}$ and onto haloes as $f^{\rm{halo}}_{\rm{acc,hot/cold}}$.

The top panel of Fig.~\ref{accretion_hotcold_plot} shows the median fraction of gas accretion onto galaxies, $f^{\rm{galaxy}}_{\rm{acc,hot}}$, as a function of halo mass for different redshifts. The fractions were taken from the Ref-L100N1504 simulation and the error bars in the figure show the $1\sigma$ scatter. We find that in $>10^{12.8}\Msun$ haloes at $z=0$ ($>10^{13.3}\Msun$ haloes at $z=1$), $\dot{M}_{\rm{gas,galaxy}}$ changes from being cold-mode dominated to hot-mode dominated, and at fixed halo mass $f^{\rm{galaxy}}_{\rm{acc,hot}}$ decreases with increasing redshift. This is expected, since at low redshift there are fewer cold filaments penetrating the hot halo and delivering cold gas within galaxies (see Paper I, Section 3). 

In Paper I we compared two different methods to select particles accreted hot or cold based on the maximum temperature ($T_{\rm{max}}$) ever reached by the gas particle and the temperature ($T_{\rm{post-shock}}$) of the gas particle after being accreted. By applying the $T_{\rm{max}}$ method, which is the most commonly used (e.g. \citealt{Keres05,Keres09,Faucher,vandeVoort11,Nelson13}), we find that the fraction of hot accretion onto haloes is in very good agreement with \citet{vandeVoort11}, as it decreases with increasing redshift at fixed halo mass (see Paper I for a detailed comparison). However, the hot accretion fraction onto galaxies calculated in this work appears to deviate strongly from previous works that claimed that there is almost no cold accretion onto galaxies at $z\sim 2$. For example, \citet{Nelson15a} (as well as \citealt{Keres09}) found that at $z=2$ cold accretion of external diffuse gas accounts for only $10\%$($30\%$) of the total accretion onto central galaxies of $10^{12}\Msun$ haloes without (with) AGN/stellar feedback. \citet{Nelson15a} used simulations run with the AREPO code (Springel 2010) and defined gas to be in the hot mode of accretion if the maximum past temperature of the gas was larger than the virial temperature of the host halo at the accretion time (time of the most recent radial crossing). In Paper I, as well as throughout this work, we apply the $T_{\rm{post-shock}}$ method, based on the temperature of the gas particle after accretion, to calculate the hot/cold modes of gas accretion. We find that cold accretion onto galaxies in $10^{12}\Msun$ haloes accounts for $50\%$ ($70\%$) of the total at $z=0$ ($z=2$) using the $T_{\rm{max}}$ criteria, but it accounts for $95\%$ ($98\%$) using the $T_{\rm{post-shock}}$ criteria.

We believe that $T_{\rm{max}}$ is less suitable for identifying cold gas accretion for the following reasons. Firstly, it is possible for gas to go through a shock but immediately cool afterwards. In this case if gas is mostly cold except at a point in space and for a short period of time, numerical studies using $T_{\rm{max}}$ would label it as hot accretion but observations would indicate a cold flow. Secondly, outflows can heat the surrounding gas particles, which can reach high temperatures while being expelled from the galaxy. Such particles eventually cool and are re-accreted onto the galaxy. However, even if they do so via the cold mode they are classified as hot mode accretion by the $T_{\rm{max}}$ criterion (for more details see Paper I). 

The bottom panel of Fig.~\ref{accretion_hotcold_plot} shows a comparison between the fraction of gas accreted hot onto haloes (solid line) and onto galaxies (dashed line) at $z=0$. It can be seen that while $70\%$ of the gas crosses $R_{200}$ in hot mode for $10^{12}\Msun$ haloes, less than $5\%$ crosses $0.15\times R_{200}$ in hot mode and reaches the galaxy. However this changes in higher mass haloes. In $10^{13}\Msun$ haloes, for instance, while $98\%$ of the gas crosses $R_{200}$ in hot mode, $80\%$ crosses $0.15\times R_{200}$ in hot mode. The increasing amount of warm low-density gas that reaches the galaxies in high-mass haloes seems to indicate that while the hot halo forms in $10^{11.7}\Msun$ haloes (see Paper I for details), the cooling flow from the hot halo develops in haloes with masses between $10^{12}-10^{13}\Msun$. 

Note that a change in the temperature threshold not only modifies the fraction of hot mode accretion, but also the mass-scale at which the hot halo forms. In Paper I we develop a semianalytic model to estimate a `critical halo mass' for hot halo formation that depends on the build up of the hot gas mass in the halo as well as on $f_{\rm{acc,hot}}$. We show that changing $f_{\rm{acc,hot}}$ from 1 to 0.5 increases the mass-scale of hot halo formation from $10^{11.4}$ to $10^{11.7}\Msun$, respectively (see Section 5.3.1 and Fig. 12 of Paper I for further details). In the following sections we show that the halo mass at which the hot halo cooling flow develops depends strongly on AGN feedback but not on stellar feedback.

\subsection{Impact of stellar feedback}\label{stellar_fb}

It has been shown that the inflow rate of gas onto galaxies sensitively depends not only on definition (as discussed in Section~\ref{Methodology}), but also on feedback physics (e.g. \citealt{Oppenheimer10,vandeVoort11,Faucher,Nelson15a,Ubler14}). Recently, \citet{Nelson15a} compared two simulations run with the AREPO code. While one included energetic feedback from star formation driven winds as well as supermassive black holes, the other did not include any treatment of metal line cooling, stellar or black hole feedback. They found that feedback strongly suppresses the {\it{net}} accretion rate onto central galaxies (counted as the number of gas tracers crossing the radius $0.15R_{200}$): by a factor of $\sim$3 at $z=5$, and a factor of $\sim$10 at $z=1$. A similar conclusion was reached by \citet{vandeVoort11}, who showed that the effects of stellar feedback and metal-line cooling are much stronger for accretion onto galaxies than for accretion onto haloes, and can result in differences of an order of magnitude. 

\begin{figure} 
  \centering
  \vspace{-0.3cm}
  \includegraphics[angle=0,width=0.49\textwidth]{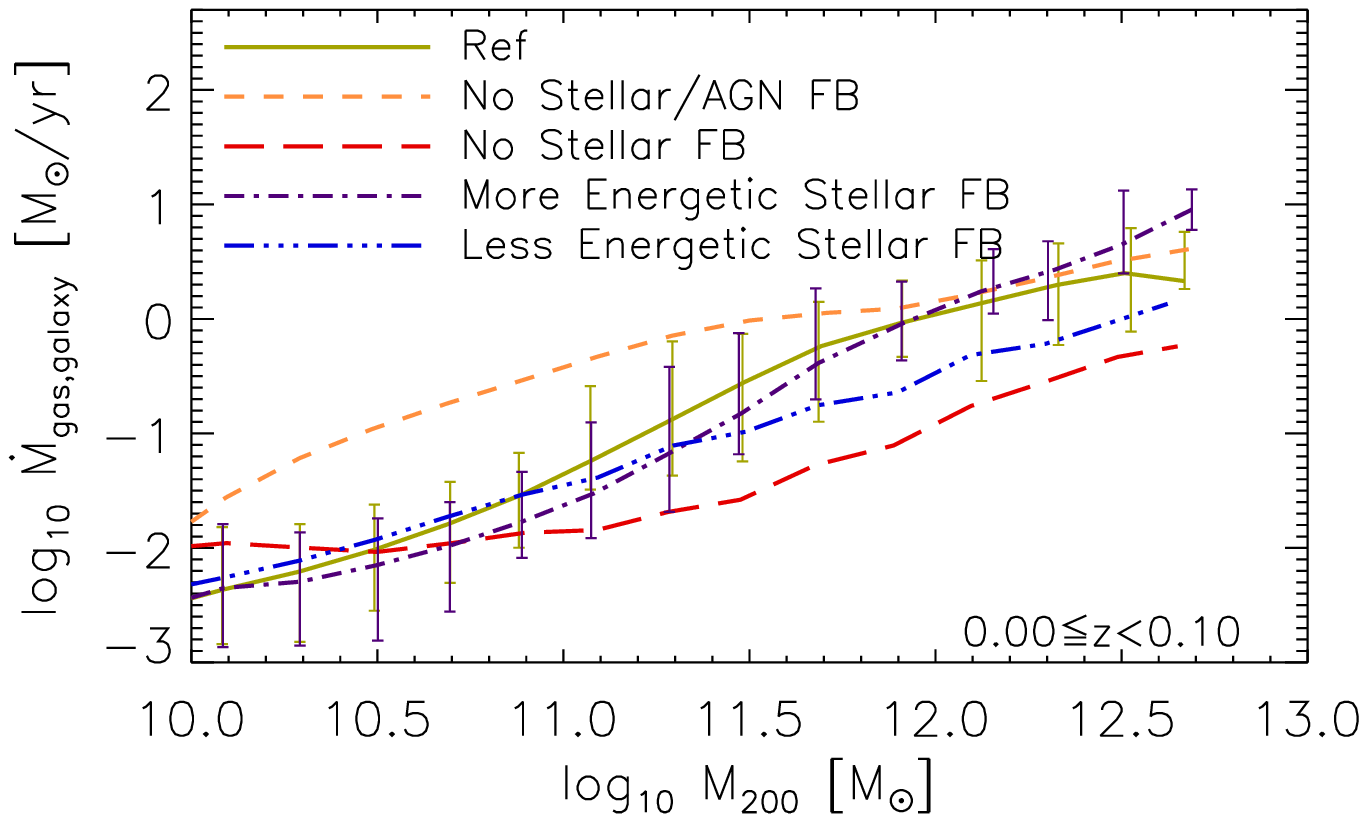}\\
  \vspace{-0.4cm}
  \includegraphics[angle=0,width=0.49\textwidth]{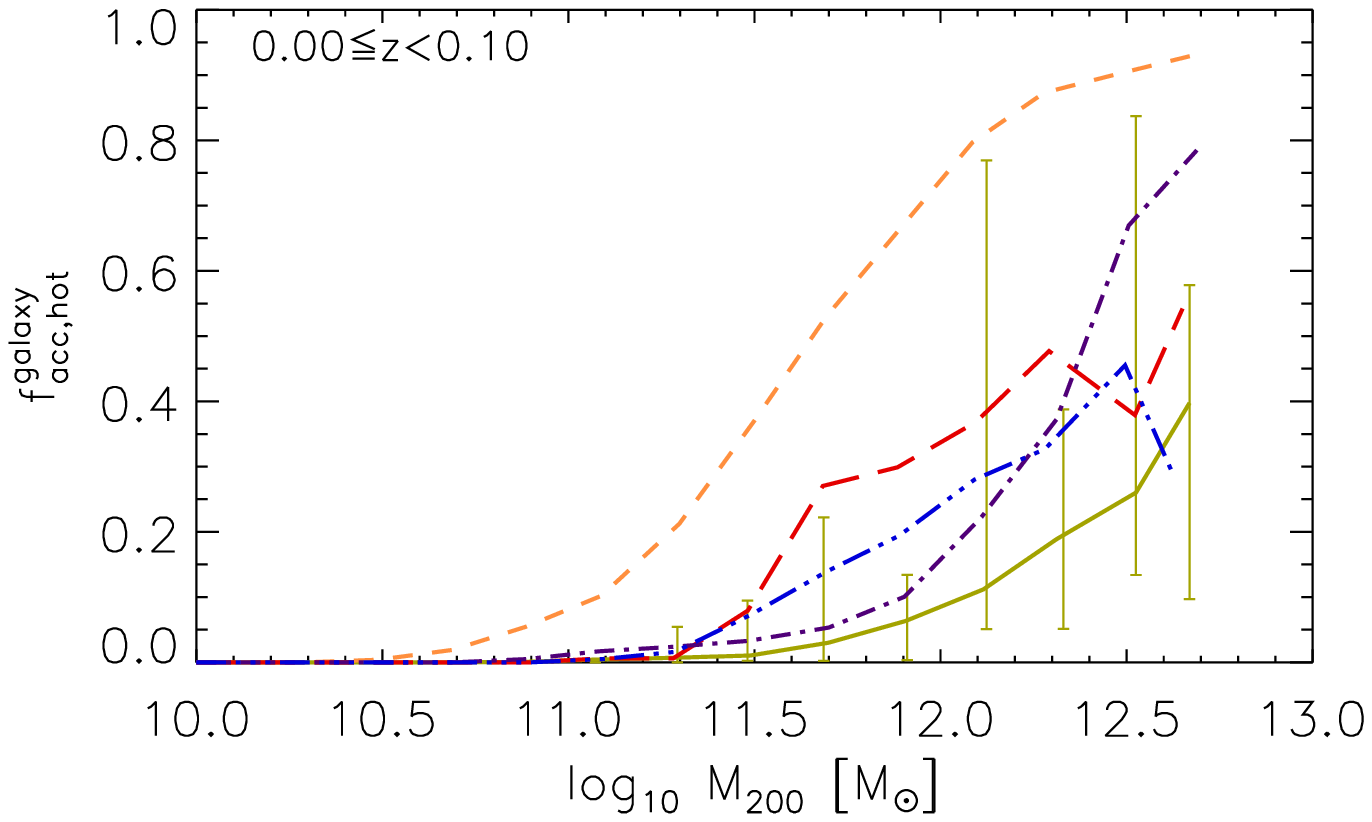}\\
%  \vspace{-0.4cm}
 % \includegraphics[angle=0,width=0.45\textwidth]{./plots/Gal_accretion_HotFrac_stellar_fb.ps}
%  \vspace{-0.2cm}
  \caption{Top panel: Gas accretion rate onto central galaxies as a function of halo mass. Bottom panel: fraction of hot mode gas accretion onto central galaxies as a function of halo mass in the redshift range $0\leq z<0.1$. The panels compare the median accretion rates/hot fractions from simulations with the same resolution (L025N0376 box), but with varying feedback models. These include standard stellar feedback (Ref, solid green line), more energetic stellar feedback (purple dot-dashed line), less energetic stellar feedback (blue dot-dashed line), no stellar feedback (red long-dashed line), and no stellar/AGN feedback (orange dashed line). The error bars show the 16-84th percentiles. Depending on the halo mass, changes in the efficiency of stellar feedback can either increase or decrease the accretion rates onto galaxies.}% The trends can plausibly be explained by considering the re-accretion of gas ejected by progenitor galaxies and by the suppression of black hole growth, and hence AGN feedback, by stellar feedback.}
\label{gal_accr_stellar_fb}
\end{figure}

In scenarios with energetic stellar feedback, the net galaxy accretion rates can be higher due to `recycling accretion'. Stellar driven winds blow gas out of the galaxy, but not out of the halo, as a result the same gas elements are accreted onto the galaxy multiple times (\citealt{Oppenheimer10}, see also \citealt{vandeVoort17b} for a recent review). \citet{Ubler14} implemented a hybrid thermal/kinetic stellar feedback scheme, and calculated the gas accretion histories onto discs as a function of cosmic time. They found that the amount of re-accreted gas can be a factor of 10 larger in the strong feedback models with respect to the weak feedback models, and tends to dominate the net accretion at $z<1$. 

To investigate the effect of stellar feedback on the gas accretion rate onto galaxies from the EAGLE simulations, we compare the reference model with identical resolution simulations (the L025N0376 model) that include more/less energetic stellar feedback, no stellar feedback and no stellar/no AGN feedback (referred to as More/Less Energetic Stellar FB, No Stellar FB and No Stellar/AGN FB, respectively). The Ref, More/Less Energetic Stellar FB and No Stellar FB simulations employ the same feedback prescription and choice of parameters for AGN feedback, but the energy budget expelled by the stellar feedback is varied, and in the last case (No Stellar FB) stellar feedback is switched off. In the case of the No Stellar/AGN FB simulation, both stellar and AGN feedback are switched off.

In the EAGLE simulations, the probability for a neighboring SPH particle to be heated by stellar outflows is determined by the fraction of the energy budget available for feedback, $f_{\rm{th}}$, which is defined as $f_{\rm{th}}$=$f_{\rm{th,min}}+f(Z,n_{\rm{H}})(f_{\rm{th,max}}-f_{\rm{th,min}})$ (with $f_{\rm{th,max}}$, $f_{\rm{th,min}}$ the maximum and minimum asymptotic values and $f(Z,n_{\rm{H}})$ a function of the gas metallicity and density). In the Ref model $f_{\rm{th,max}}=3.0$ and $f_{\rm{th,min}}=0.3$, whereas in the More/Less Energetic FB models the thresholds are $f_{\rm{th,max}}=6.0$ and $f_{\rm{th,min}}=0.6$ for the More Energetic Stellar FB case, and $f_{\rm{th,max}}=1.5$ and $f_{\rm{th,min}}=0.15$ for the Less Energetic Stellar FB case, respectively. This does not mean that stellar driven winds are stronger/weaker or blow more/less gas out of the galaxy, but rather that in the More/Less Energetic Stellar FB model the energy injected per unit mass of stars formed is twice/half of the amount used in the Ref model.

The top panel of Fig.~\ref{gal_accr_stellar_fb} shows the gas accretion rate onto galaxies at the centers of dark matter haloes, as a function of halo mass in the redshift range $0\leq z < 0.1$. From the panel it can be seen that $\dot{M}_{\rm{gas,galaxy}}$ from the More Energetic Stellar FB model is (on average) a factor of 2 lower than $\dot{M}_{\rm{gas,galaxy}}$ from Ref, but it increases for galaxies in haloes larger than $10^{12}\Msun$ (by up to a factor 4 larger than Ref for galaxies in $10^{12.7}\Msun$ haloes), suggesting that at these halo masses galaxies re-accrete gas that was ejected by stellar feedback in lower-mass progenitors. When stellar feedback is half as energetic, $\dot{M}_{\rm{gas,galaxy}}$ decreases by up to 0.6 dex compared to Ref for galaxies in $\geq 10^{11}\Msun$ haloes. A possible reason for this difference is a lower rate of re-accreted gas or a more efficient AGN feedback at lower halo masses. The latter is also due to the less energetic stellar feedback. \citet{Bower17} showed that in EAGLE stellar feedback limits BH growth in low-mass haloes. If stellar feedback is half as energetic the central BH is able to start growing earlier, thus producing more efficient AGN feedback.

The panel also shows that when stellar feedback is switched off, AGN feedback affects the gas accretion rates onto galaxies residing in low-mass haloes. We find that for galaxies in $\geq 10^{10.5}\Msun$ haloes, $\dot{M}_{\rm{gas,galaxy}}$ decreases by up to 1.5 dex relative to Ref. When there is no stellar feedback, the central black hole in low-mass galaxies is able to grow by more than 1 order of magnitude with respect to the Ref model (\citealt{Bower17}). Thus the outflows expelled by a much more massive black hole suppress the gas infall rates in low-mass galaxies. Indeed, when both stellar and AGN feedback are turned off, there is no mechanism that prevents gas from cooling. Therefore the rates of gas accretion are higher by up one dex than in Ref. 

We also analyse whether the trends described in Fig.~\ref{gal_accr_stellar_fb} depend on redshift. We obtain that they do not, at $z=2$ the trends are consistent with the $z=0$ results.
%We obtain that at $z=2$ the accretion rate for the more energetic stellar feedback case (or the no stellar feedback case) is lower than for Ref, in agreement with our results at $z=0$.

The bottom panel of Fig.~\ref{gal_accr_stellar_fb} shows the fraction of gas accreted in the hot mode onto central galaxies as a function of halo mass. In the models, stellar feedback can either increase or decrease $f_{\rm{acc,hot}}$ by generating winds that heat the gas (before or after accretion). We find that the trend of hot fraction with feedback variation is very complex and difficult to predict. Interestingly, it is nonetheless the No Stellar/AGN FB model, that has the highest hot fractions, by up to an order of magnitude larger than the Ref model. This suggests that feedback preferentially prevents hot gas from reaching the galaxy. This is expected, because energy-driven winds will take the path of least resistance, thus avoiding the cold streams (e.g. \citealt{Theuns02}). 

Note that in this work, the cold/hot temperature cut we used to separate cold from hot accretion is applied after accretion onto the galaxy. Therefore, it could happen that the hot fraction artificially increases due to heating by stellar feedback, even though Fig.~\ref{gal_accr_stellar_fb} seems to indicate otherwise. 

\begin{figure} 
  \centering
  \vspace{-0.3cm}
  \includegraphics[angle=0,width=0.49\textwidth]{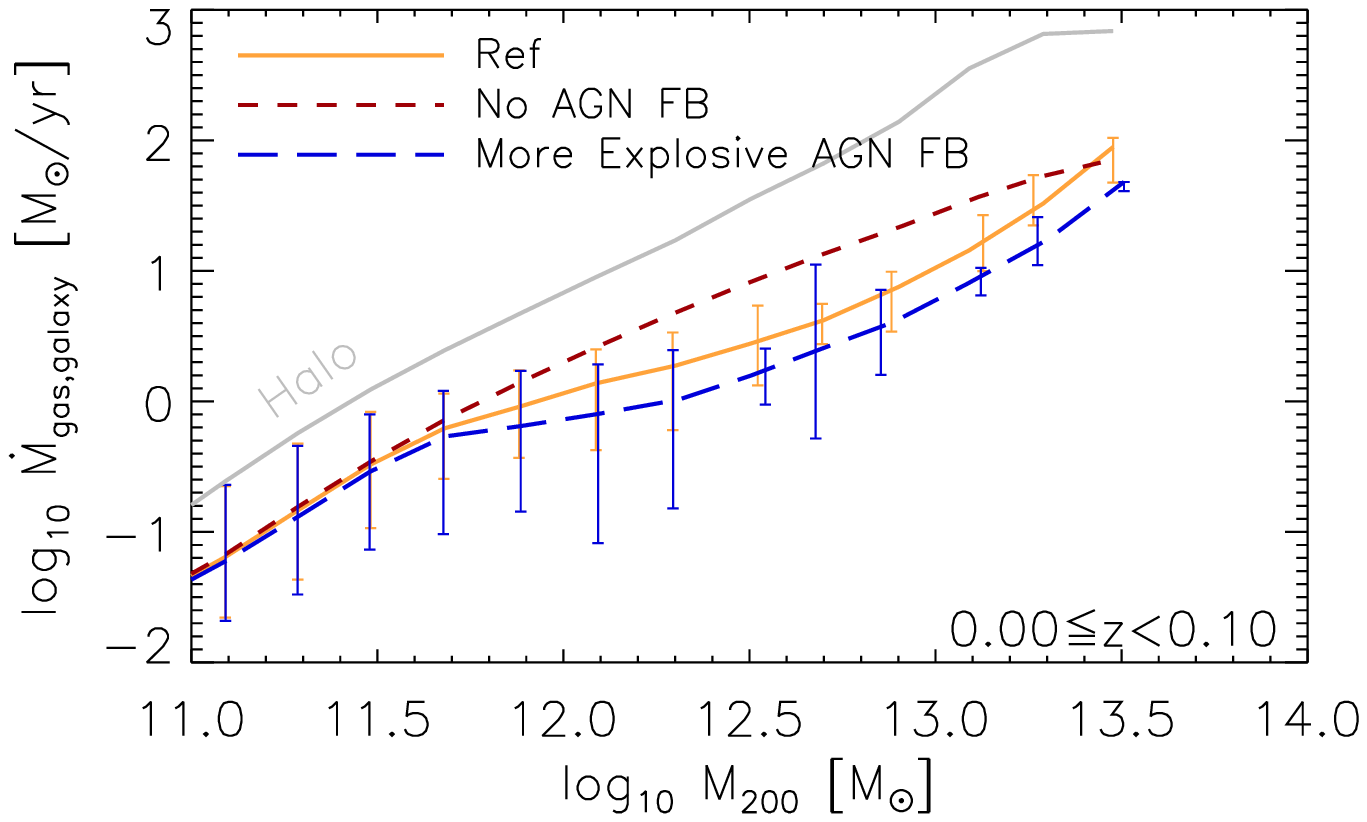}\\
  \vspace{-0.4cm}
  \includegraphics[angle=0,width=0.49\textwidth]{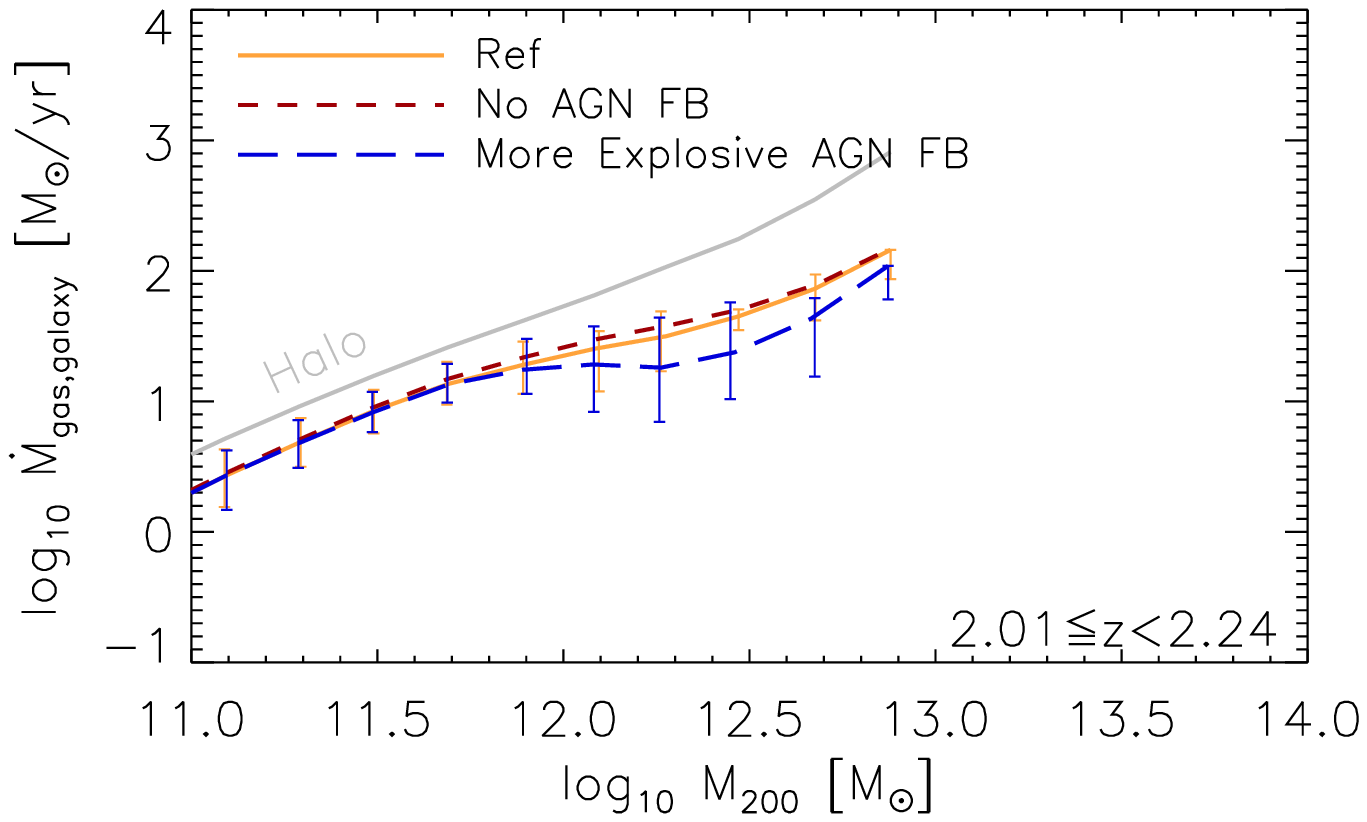}\\
  \vspace{-0.4cm}
  \includegraphics[angle=0,width=0.49\textwidth]{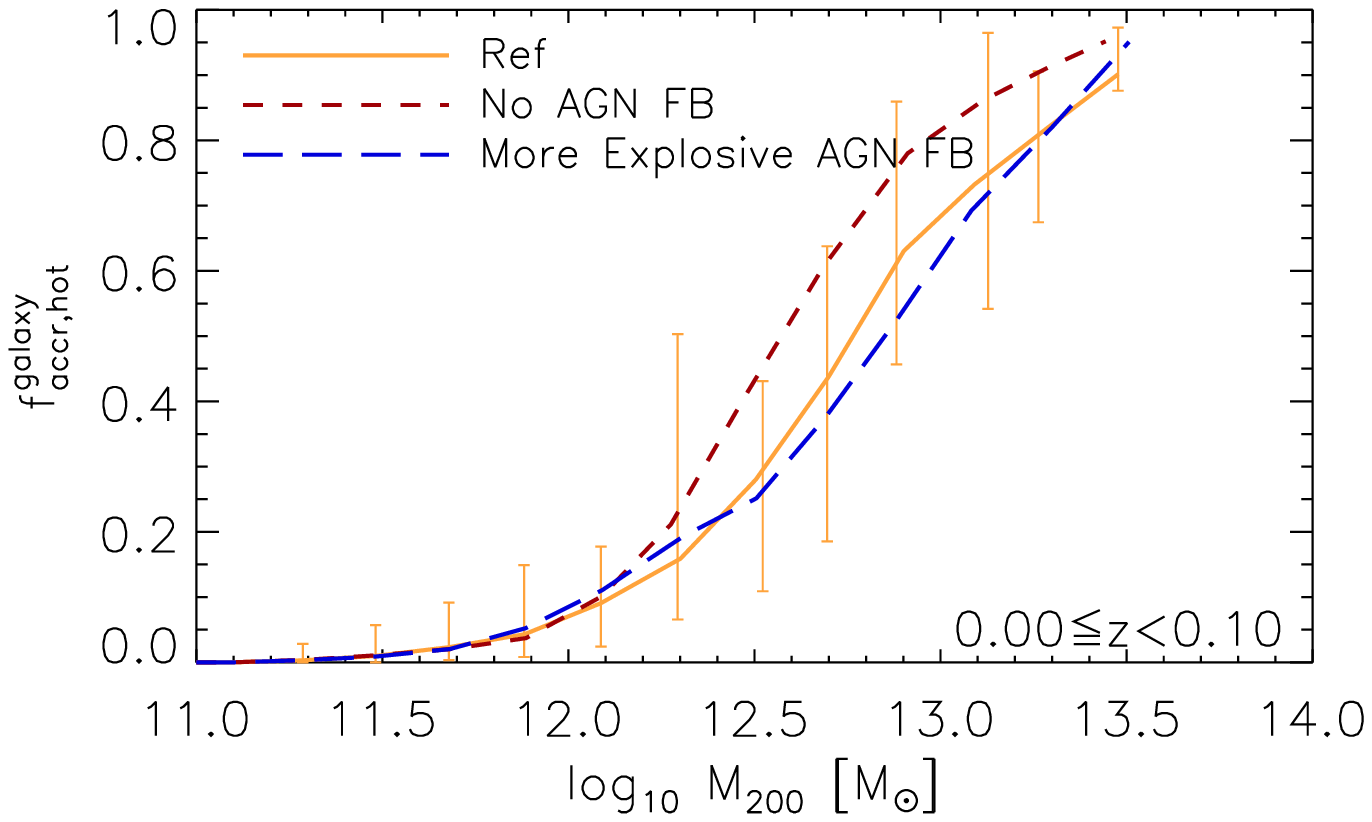}
%  \vspace{-0.4cm}
%  \includegraphics[angle=0,width=0.45\textwidth]{./plots/Gal_accretion_HotFrac_agn_fb.ps}
%  \vspace{-0.2cm}
  \caption{Median gas accretion rate onto central galaxies as a function of halo mass in the redshift range $0\leq z<0.1$ (top panel) and $2\leq z<2.2$ (middle panel). The top and middle panels compare accretion rates from same-resolution simulations (L050N0752) but with standard AGN feedback (Ref, solid orange line), No AGN feedback (red short-dashed line) and more explosive AGN feedback (blue long-dashed line). The solid grey lines correspond to median gas accretion rate onto haloes as a function of halo mass in the redshift range $0\leq z<0.1$ (top panel) and $2\leq z<2.2$ (middle panel). The bottom panel compares the same simulations, but shows the median fraction of gas accreted onto galaxies in the hot mode as a function of halo mass at $0\leq z<0.1$. Error bars show the 16-84th percentiles. When AGN feedback is switched off, $\dot{M}_{\rm{gas,galaxy}}$ does not flatten at ${\sim}10^{12}\Msun$ and increases with halo mass at the same rate as the gas accretion onto haloes but with a lower normalization. This shows that AGN feedback is the mechanism responsible for preventing hot gas from cooling and falling onto central galaxies in massive haloes.}
\label{gal_accr_agn_fb}
\end{figure}

\subsection{Impact of AGN feedback}\label{AGN_fb}

To understand how AGN feedback alters $\dot{M}_{\rm{gas,galaxy}}$ in massive haloes, we compare simulations that include the same stellar feedback scheme but different prescription for AGN feedback, varying from no AGN feedback (No AGN FB), standard AGN feedback (Ref model), to more explosive AGN feedback (More Explosive AGN FB). In the EAGLE simulations, the difference between the Ref and More Explosive AGN FB simulations is the temperature increment of stochastic AGN heating ($\Delta T_{\rm{AGN}}$), which is $\Delta T_{\rm{AGN}}=10^{8.5}$ K in the Ref model and $\Delta T_{\rm{AGN}}=10^{9.0}$ K in the More Explosive AGN FB model. This means that the AGN feedback is more explosive and intermittent, but the energy injected per unit mass accreted by the BH does not change with respect to the Ref model.

Fig.~\ref{gal_accr_agn_fb} shows $\dot{M}_{\rm{gas,galaxy}}(M_{200})$ for the different models in the redshift range $0\leq z < 0.1$ (top panel) and $2\leq z < 2.24$ (middle panel). For comparison, the panels also show the gas accretion rate onto haloes (i.e. within $R_{200}$ as taken from the Ref model) as grey solid lines. Note that the More Explosive AGN FB and No AGN FB models were run in 50 Mpc volumes, so those do not contain haloes more massive than $10^{13.5}\Msun$. The top panel shows that at $z=0$ AGN feedback suppresses $\dot{M}_{\rm{gas,galaxy}}$ in massive haloes ($>10^{12}\Msun$) by up to 0.6 dex in the Ref model, and up to 0.8 dex in the More Explosive AGN FB model. In the No AGN FB simulation, $\dot{M}_{\rm{gas,galaxy}}$ does not flatten at $\sim10^{12}\Msun$ and increases with halo mass at the same rate as the gas accretion onto haloes but with a 0.5 dex lower normalization. This indicates that AGN feedback is the mechanism responsible for preventing hot gas from cooling and falling onto the central galaxies. However, the situation differs at $\approx 2$. In this case the middle panel shows that $\dot{M}_{\rm{gas,galaxy}}$ flattens with increasing halo mass for $10^{12}-10^{12.5}\Msun$ haloes in both the Ref and the No AGN FB model. For the More Explosive AGN FB model the flattening is however more pronounced.

We believe that the flattening of $\dot{M}_{\rm{gas,galaxy}}$ in massive haloes (${>}10^{12}\Msun$) at $z=0$ can be explained by the rate of gas cooling from the hot halo. A hot hydrostatic atmosphere is formed in ${\sim}10^{11.7}\Msun$ haloes (Paper I) as a result of heating by accretion shocks. Some time after the hot halo is formed, gas begins to cool and fall onto the central galaxy, but it can also be reheated or be prevented from accreting by AGN feedback. When there is no AGN feedback preventing the shock-heated gas in the halo from cooling, the hot gas is able to cool over a short time-scale. As a result a larger amount of gas cools from the hot halo raising $\dot{M}_{\rm{gas,galaxy}}$ in the No AGN FB model with respect to the Ref model. This can also be seen from the bottom panel of Fig.~\ref{gal_accr_agn_fb}, which shows the fraction of gas accreted onto galaxies in the hot mode as a function of halo mass for the different simulations. We find that the hot fraction does not depend strongly on the explosiveness of AGN feedback, but it does increase if AGN feedback is turned off. This indicates that a larger fraction of gas cooling from the hot halo is able to reach the galaxy without AGN feedback.

%We believe that the flattening of $\dot{M}_{\rm{gas,galaxy}}$ in massive haloes ($>10^{12}\Msun$) at $z=0$ can be explained by the rate of gas cooling from the hot halo, that it is developed some time after the hot corona is formed (in $\sim 10^{11.7}\Msun$ haloes at $z=0$, Paper I) and it is further suppressed with AGN feedback. 

We further explore the validity of this hypothesis in the following section, where we develop a semi-analytic model of the gas accretion rate onto galaxies that includes the heating and cooling rates of gas from the hot halo.

%\subsection{Impact of stellar and AGN feedback on accretion rates onto haloes}\label{halo_accretion}

%In this section we compare the gas accretion rates onto haloes, $\dot{M}_{\rm{gas,halo}}$, taken from simulations with different feedback prescriptions (i.e. more explosive AGN, more/less energetic stellar feedback). Fig.~\ref{accretion_halos}
% the rates are in close agreement with each other (as also shown by \citealt{vandeVoort11}). The largest difference is in $10^{12}\Msun$ haloes, where a more explosive AGN feedback decreases $\dot{M}_{\rm{gas,halo}}$ in 0.2 dex with respect to Ref and the lack of AGN feedback increases $\dot{M}_{\rm{gas,halo}}$ in 0.1 dex. In the case of stellar feedback, more energetic feedback increases $\dot{M}_{\rm{gas,halo}}$ by 0.1 dex with respect to Ref for $10^{12}\Msun$ haloes, whereas less energetic feedback decreases $\dot{M}_{\rm{gas,halo}}$ by up to 0.4 dex. In the no feedback simulations, the $\dot{M}_{\rm{gas,halo}}$ median values are in perfect agreement with \citet{Correa15c}. 

\section{The hot halo cooling flow}\label{AGN_hot_halo_sec}

In this section we show that the increase in $\dot{M}_{\rm{gas,galaxy}}$ with respect to $\dot{M}_{\rm{ISM}}$ in haloes larger than $10^{12.5}\Msun$ (see Fig. 1) can be explained by the rate of gas cooling from the hot halo. To do so, we develop a model of gas accretion onto galaxies that depends on the hot/cold fractions of gas accretion onto haloes, and on the shock-heating and cooling rates of gas from the hot halo. We briefly describe the model in the following subsection and in Section~\ref{results} we compare the result of the model with the simulation output.

\subsection{Semi-analytic model of gas accretion onto galaxies}\label{model_section}

In Section \ref{hot_cold_modes_of_accretion} we showed that the gas accretion onto galaxies can be decomposed into the sum of two modes of accretion, hot and cold. We consider these two modes in our model and calculate $\dot{M}_{\rm{gas,galaxy}}$ in terms of the rate of gas cooling from the hot halo, $\dot{M}_{\rm{cooling}}$, and the rate of cold gas accretion onto haloes in the form of filaments, $f^{\rm{halo}}_{\rm{acc,cold}}\dot{M}_{\rm{gas,halo}}$, as

\begin{eqnarray}\label{Mgalaxy1}
\dot{M}_{\rm{gas,galaxy}} &\propto & \dot{M}_{\rm{cooling}}+f^{\rm{halo}}_{\rm{acc,cold}}\dot{M}_{\rm{gas,halo}}.
\end{eqnarray}

\noindent Here $\dot{M}_{\rm{gas,halo}}$ is the gas accretion rate onto haloes in the presence of feedback (given by eqs.~\ref{halo_accr1}-\ref{halo_accr8}), and $f^{\rm{halo}}_{\rm{acc,hot/cold}}(M_{200},z)$ are the hot/cold fraction of gas accretion onto haloes. In eq.~(\ref{Mgalaxy1}) we assume that the cold accretion onto the galaxy is directly proportional to the cold accretion onto the halo, with the latter given by the following best-fit expressions from Paper I

\begin{eqnarray}\label{fhot1}
f^{\rm{halo}}_{\rm{acc,hot}}(x) &=& [{\rm{exp}}(-4.3[x+0.15])+1]^{-1},\\\label{fhot2}
f^{\rm{halo}}_{\rm{acc,cold}}(x) &=& 1-f^{\rm{halo}}_{\rm{acc,hot}}(x),\\\label{fhot3}
x &=& \log_{10}(M_{200}/10^{12}\Msun).
\end{eqnarray}

We define $M_{\rm{cooling}}$ as the hot gas mass in the halo contained within the cooling radius, $r_{\rm{cool}}$ (with $r_{\rm{cool}}\leq R_{200}$). Therefore $M_{\rm{cooling}}=M_{\rm{hot}}\frac{r_{\rm{cool}}}{R_{200}}$, where $M_{\rm{hot}}$ is the total hot gas mass in the halo and we assumed an isothermal profile. We then assume that the variation of $M_{\rm{cooling}}$ in time is mainly driven by the variation of $M_{\rm{hot}}$ and obtain

%$r_{\rm{cool}}/R_{200}$ remains roughly constant during a short time step and $\dot{M}_{\rm{cool}}$ results

\begin{eqnarray}\label{xi0}
\dot{M}_{\rm{cooling}} &\approx & \dot{M}_{\rm{hot}}\frac{r_{\rm{cool}}}{R_{200}},\\\label{xi1}
\dot{M}_{\rm{cooling}} &\approx & f^{\rm{halo}}_{\rm{acc,hot}}\dot{M}_{\rm{gas,halo}}\frac{r_{\rm{cool}}}{R_{200}},
\end{eqnarray}

\noindent In eq.~(\ref{xi0}) we assumed that the time scale over which $M_{\rm{cooling}}$ varies is short enough for the halo not to grow in mass and for $\dot{r}_{\rm{cool}}=\dot{R}_{200}=0$. Note that for this equation we assume an isothermal density profile for simplicity, but to better model the flow of gas onto the galaxy we use the actual density profile in the calculation of the cooling rate.

In eq.~(\ref{xi1}) we assumed that $\dot{M}_{\rm{cooling}}$ is determined by the rate of replenishment from hot accretion onto the halo, $\dot{M}_{\rm{hot}}=f^{\rm{halo}}_{\rm{acc,hot}}\dot{M}_{\rm{gas,halo}}$, and therefore $\dot{f}^{\rm{halo}}_{\rm{acc,hot}}=0$. These are first order approximations accurate enough for our semi-analytic model.

Eq.~(\ref{xi1}) gives the cooling radius a new physical meaning. Besides being the radius within which all gas is able to cool, we now interpret it as the fraction of shock-heated gas that cools (${\frac{r_{\rm{cool}}}{R_{200}}=\frac{\dot{M}_{\rm{cooling}}}{f^{\rm{halo}}_{\rm{acc,hot}}\dot{M}_{\rm{gas,halo}}}}$) and reaches the galaxy. In other words, it is the rate of the hot halo cooling flow. By substituting eq.~(\ref{xi1}) into eq~(\ref{Mgalaxy1}), the accretion rate onto galaxies becomes

\begin{equation}\label{Mgalaxy2}
\dot{M}_{\rm{gas,galaxy}} = \epsilon(f^{\rm{halo}}_{\rm{acc,hot}}\frac{r_{\rm{cool}}}{R_{200}}+f^{\rm{halo}}_{\rm{acc,cold}})\dot{M}_{\rm{gas,halo}},
\end{equation}

\noindent where $\epsilon$ is a dimensionless correction factor set to be 0.3 so that $\dot{M}_{\rm{gas,galaxy}}$ agrees with the gas accretion rate of galaxies from the Ref model in $10^{11}\Msun$ haloes (when $f^{\rm{halo}}_{\rm{acc,hot}}r_{\rm{cool}}/R_{200}+f^{\rm{halo}}_{\rm{acc,cold}}\approx 1$ and $\dot{M}_{\rm{gas,galaxy}}\approx 0.3\dot{M}_{\rm{gas,halo}}$). Note that $\epsilon$ captures the suppression of accretion onto galaxies due to intrahalo feedback processes that likely depend on simulation.

The semi-analytic model given by eq.~(\ref{Mgalaxy2}) uses as input the gas accretion onto haloes ($\dot{M}_{\rm{gas,halo}}$, given by eqs.~\ref{halo_accr1}-\ref{halo_accr5}) and the hot/cold fraction of gas accretion onto haloes ($f^{\rm{halo}}_{\rm{acc,hot/cold}}$, given by eqs.~\ref{fhot1}-\ref{fhot3}). To compute the model we also need the ratio $r_{\rm{cool}}/R_{200}$ as a function of halo mass and redshift. We calculate it by equating the heating ($\Gamma_{\rm{heat}}$) and cooling ($\Gamma_{\rm{cool}}$) rates (energy per unit time) of gas from the hot halo (derived in Paper I) 

\begin{eqnarray}\label{gamma_heat}
\Gamma_{\rm{heat}}(M_{200})= &\frac{3}{2}\frac{k_{\rm{B}}T_{\rm{vir}}}{\mu m_{\rm{p}}}\frac{\Omega_{\rm{b}}}{\Omega_{\rm{m}}}\dot{M}_{200}\left[\frac{2}{3}f_{\rm{hot}}+f^{\rm{halo}}_{\rm{acc,hot}}\right],&\\\label{gamma_cool}
\Gamma_{\rmn{cool}}(M_{200},r)= &M_{\rm{hot}}\frac{\Lambda[T_{\rm{hot}},Z_{\rm{hot}},\rho_{\rm{hot-gas}}(r)]}{\rho_{\rm{hot-gas}}(r)}.&
\end{eqnarray}

\noindent In eqs. (\ref{gamma_heat}) and (\ref{gamma_cool}), $\rho_{\rm{hot}}$, $T_{\rm{hot}}$ and $Z_{\rm{hot}}$ are the characteristic hot gas density, temperature and metallicity respectively, and $f_{\rm{hot}}$ is the fraction of gas in the halo that is hot ($f_{\rm{hot}}\equiv M_{\rm{hot}}/[\frac{\Omega_{\rm{b}}}{\Omega_{\rm{m}}}M_{200}]$), which is given by the following best-fit relation from Paper I

\begin{eqnarray}\label{Mhot2}
\log_{10}\left(\frac{M_{\rm{hot}}}{(\frac{\Omega_{\rm{b}}}{\Omega_{\rm{m}}})M_{200}}\right)&=& -0.8+0.5x-0.05x^{2},\\\nonumber
x &=& \log_{10}(M_{200}/10^{12}\Msun).
\end{eqnarray}

\noindent Also, in eqs. (\ref{gamma_heat}) and (\ref{gamma_cool}), $\dot{M}_{200}$ is the dark matter accretion rate of the halo and $\Lambda$ is the net cooling rate per unit volume. The only parameter that depends on radius is $\rho_{\rm{hot-gas}}(r)$, which we estimate by extracting the hot gas density profiles from the simulation and interpolating to obtain the gas density as a function of radius and halo mass (see Appendix~\ref{density_profile} for details of the density profiles). 

To compute $r_{\rm{cool}}/R_{200}$, we assume that $T_{\rm{hot}}$ is equal to the halo virial temperature, $Z_{\rm{hot}}=0.1Z_{\odot}$ and obtain $\Lambda$ from the tabulated cooling rates given by \citet{Wiersma09a}. In Paper I we explored the relation between the mass-weighted median metallicity of the hot gas with halo mass and redshift, and found that $Z_{\rm{hot}}\sim 0.1Z_{\odot}$ for $10^{12}\Msun$ haloes at $z=0$ and varies by up to a factor of 2.5 in the halo mass range $10^{11}-10^{14}\Msun$. We find that changing the metallicity with halo mass, changes the normalization of the relation $r_{\rm{cool}}/R_{200}-M_{200}$, but the qualitative result remains the same. Finally we obtain $\dot{M}_{200}$ from the analytic model of \citet{Correa15c}. Note that the various best-fit expressions presented in this section were derived in Paper I by fitting to the results of the Ref model, therefore they depend on simulations. For further details of the calculation of $\Gamma_{\rm{heat}}$ and $\Gamma_{\rm{cool}}$ see Paper I. In the following section we compare the result of the semi-analytic model with the simulation output for $z=0$ only, but we have found that the model works well in the regime $z=0-4$.

\begin{figure} 
  \centering
  \subfloat{\includegraphics[angle=0,width=0.49\textwidth]{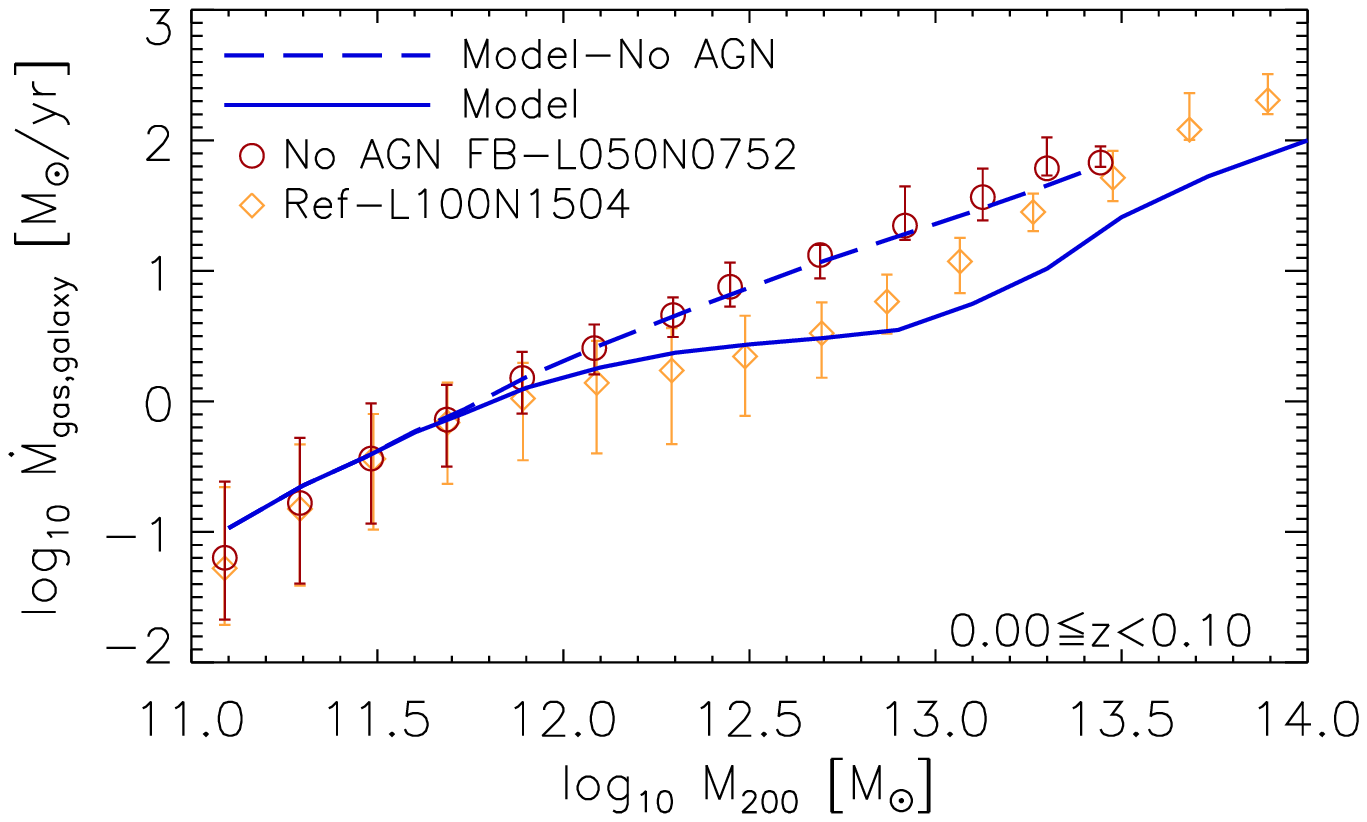}}\\
  \vspace{-0.68cm}
    \subfloat{\includegraphics[angle=0,width=0.49\textwidth]{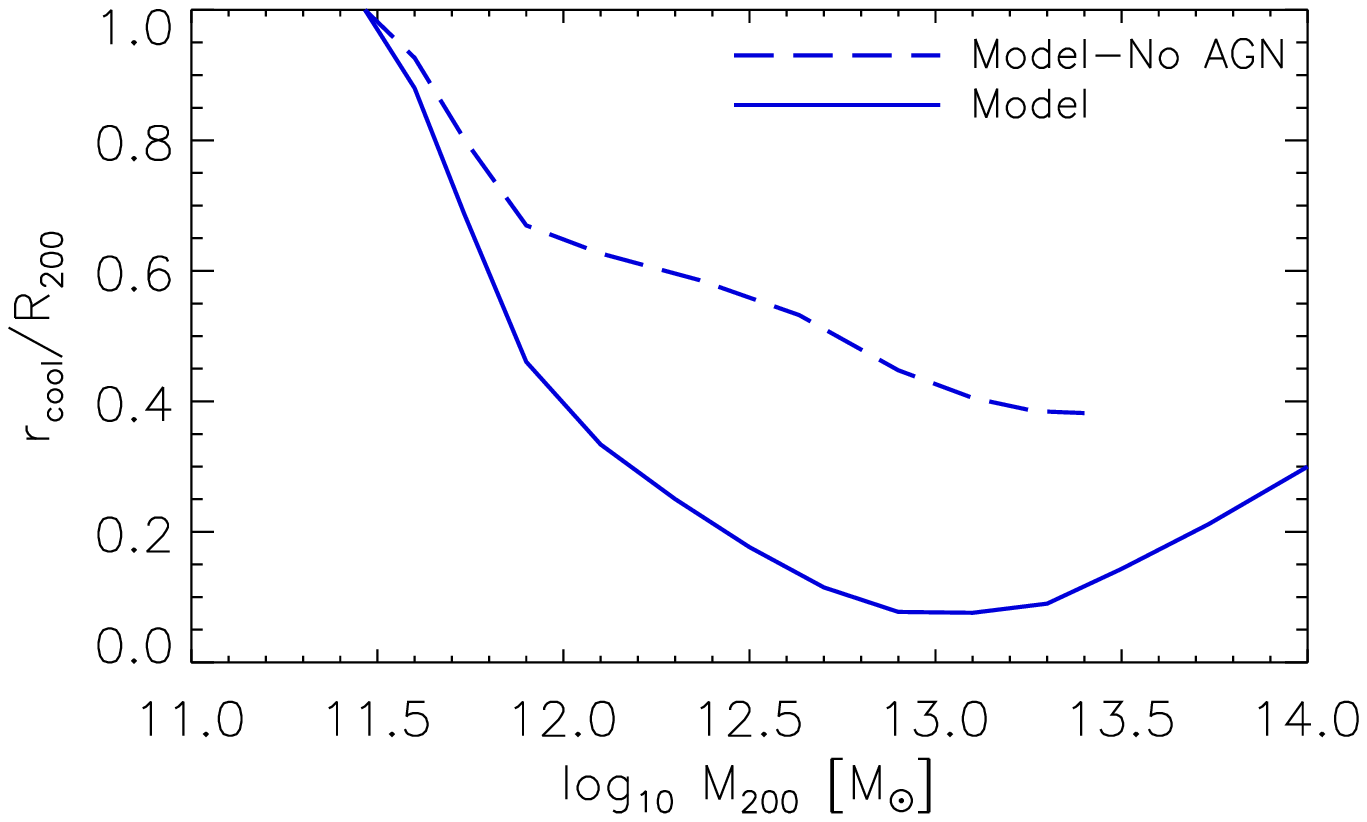}}\\
  \vspace{-0.1cm}
  \caption{Top panel: comparison between the gas accretion rates onto galaxies derived by the semi-analytic model (blue lines) and the simulation output (symbols). The orange diamond symbols correspond to the galaxy gas accretion rate taken from the Ref-L100N1504 simulation and the error bars correspond to the $1\sigma$ scatter. The red circle symbols correspond to the gas rates from the No AGN FB-L050N0752 simulation. The blue solid line corresponds to the gas accretion model, which we calculated using the heating and cooling rates from Paper I and the density profile from Section~\ref{density_profile}. Similarly, the blue dashed line corresponds to the same model but using inputs from the No AGN FB simulations. The semi-analytic model reproduces the simulation outputs. Bottom panel: analytic estimates of the cooling radius normalized by $R_{200}$ as a function of halo mass at $z=0$ from the same models shown in the top panel. The upturn in the ratio of the cooling and virial radii may indicate that in high-mass haloes AGN feedback is not sufficiently efficient.}
 \label{plot_model}
\end{figure}

\subsection{Results}\label{results}

In this section we provide insight into the physical mechanisms that drive the gas accretion rates onto galaxies. We do so by comparing the model of galaxy gas accretion derived in the previous section with the simulation output, and analysing the model's prediction in scenarios with and without AGN feedback. We emphasize that because the semi-analytic model uses input from the simulations, its predictions are not independent. We can however use it to test our physical understanding of gas accretion onto galaxies in the simulations.

The top panel of Fig.~\ref{plot_model} shows the model's prediction with AGN feedback (blue solid line) and without AGN feedback (blue dashed line). These are compared with the accretion rates from the Ref-L100N1504 (orange diamonds) and the No AGN FB-L050N0752 simulations (red circles). While the model accurately matches the gas accretion rates onto galaxies in haloes less massive than $10^{13}\Msun$ from the Ref model, at higher halo masses it under predicts the rates by up to 0.3 dex. However, the model does reproduce the qualitative trends of the flattening for $10^{12}-10^{12.5}\Msun$ and the upturn at higher halo masses. In the case of gas accretion rates onto galaxies from the No AGN FB simulation, the model is in excellent agreement. Note that for these models, the same $\epsilon$($=0.3$) correction factor is used.

The model's result when AGN feedback is included can be explained as follows. In haloes with masses lower than $10^{11.7}\Msun$, $\frac{r_{\rm{cool}}}{R_{200}}=1$, and since $f^{\rm{halo}}_{\rm{acc,hot}}+f^{\rm{halo}}_{\rm{acc,cold}}=1$, eq.~(\ref{Mgalaxy2}) gives $\dot{M}_{\rm{gas,galaxy}} = \epsilon\dot{M}_{\rm{gas,halo}}$. In haloes with masses between $10^{11.7}$ and $10^{13}\Msun$ the hot halo forms. As a result, the cooling radius is smaller than the virial radius, yielding $f^{\rm{halo}}_{\rm{acc,hot}}r_{\rm{cool}}/R_{200}+f^{\rm{halo}}_{\rm{acc,cold}}<1$, so that $\dot{M}_{\rm{gas,galaxy}}$ increases less steeply than $\dot{M}_{\rm{gas,halo}}$, remaining almost constant with halo mass. In haloes with masses larger than $10^{13}\Msun$, $\frac{r_{\rm{cool}}}{R_{200}}$ increases with halo mass, indicating that the hot halo cooling flow becomes more prominent. 

The bottom panel of Fig. 6 shows the ratio between the cooling radius and the virial radius, $\frac{r_{\rm{cool}}}{R_{200}}(M_{200})$, as a function of halo mass for the simulations with and without AGN feedback (blue solid and dashed lines, respectively). The panel shows that the ratio $\frac{r_{\rm{cool}}}{R_{200}}(M_{200})$ does not continuously decrease towards high halo masses as is commonly thought. Mathematically the upturn in the $r_{\rm{cool}}-M_{200}$ relation can be explained by the radial slope ($\gamma=$dln$\rho_{\rm{hot-gas}}/$dln$r$) of the hot gas density profile (measured between $r=0.15\times R_{200}$ and $r=R_{200}$), which changes with halo mass. The slope is roughly -2 in $10^{11.7}\Msun$ haloes, increases to -0.86 in $10^{12.7}\Msun$ haloes, and decreases to -0.94 and -1.7 in $10^{13.1}$ and $10^{13.9}\Msun$ haloes respectively. The change in the slope of $\rho_{\rm{hot-gas}}(r)$ drives the evolution of $r_{\rm{cool}}/R_{200}$, and describes the evolution in the distribution of the hot halo gas as the halo grows in mass. 

The increase of the cooling radius towards high halo masses may not be physical but the result of a deficiency of the simulations. Haloes in the Ref model more massive than $10^{13}\Msun$ contain not only 0.2 dex higher gas mass fraction (derived from virtual X-ray emission) than observed group fractions (\citealt{Schaye14}), but also too massive brightest cluster galaxies (\citealt{Bahe17}). Since the amount of hot gas (and cooling) in the halo is sensitive to the heating temperature of AGN feedback (\citealt{LeBrun14}), the disagreement with observations indicates that AGN feedback is insufficiently efficient at high halo masses.

There is a significant change in the dependence of $\dot{M}_{\rm{gas,galaxy}}$ on halo mass when there is no AGN feedback, which seems to indicate that the hot halo does not impact the galaxy gas accretion rate. However, we find that this is not the case. When the hot halo forms, it reduces the gas mass that cools, but at a lower rate when AGN feedback is not included. To understand how $\dot{M}_{\rm{gas,galaxy}}$ changes with and without AGN feedback, we refer to our model of gas accretion. As described in the previous section, the model uses as input (1) the gas accretion rate onto haloes ($\dot{M}_{\rm{gas,halo}}$), (2) the hot/cold fraction of gas accretion onto haloes ($f^{\rm{halo}}_{\rm{acc,hot/cold}}$), (3) the total hot gas mass in the halo ($M_{\rm{hot}}$) and (4) the hot gas density profile ($\rho_{\rm{hot-gas}}$). We obtain these inputs from two simulations, Ref-L100N1504 and No AGN FB-L050N0752, and predict $\dot{M}_{\rm{gas,galaxy}}$ as a function of halo mass for both models. Note that the main differences between the No AGN FB and Ref model are $M_{\rm{hot}}$ (where $M_{\rm{hot}}$ from No AGN FB is larger than from Ref in the halo mass range $10^{11.7}-10^{14}\Msun$, by a factor of 1.4 in $10^{12}\Msun$ haloes) and $\rho_{\rm{hot-gas}}$ (where the slope flattens but no as much as in the presence of AGN feedback, reaching a maximum of dln$\rho/$dln$r=-1.14$ in $10^{12.8}\Msun$ haloes from the No AGN FB model in contrast to -0.9 from the Ref model). In Paper I we found that $f^{\rm{halo}}_{\rm{acc,hot/cold}}$ and $\dot{M}_{\rm{gas,halo}}$ are roughly insensitive to feedback (in agreement with \citealt{vandeVoort11}). 

The variable from eq.~(\ref{Mgalaxy2}) that is responsible for the excellent match between $\dot{M}_{\rm{gas,galaxy}}$(no AGN) and the simulation output is $r_{\rm{cool}}/R_{200}$, whose evolution differs strongly from the scenario where AGN feedback is on. In the case of no AGN, $r_{\rm{cool}}/R_{200}$ decreases with halo mass in the mass range $10^{11.5}-10^{13}\Msun$ (shown in the bottom panel of Fig. 6), but it does not reach the same minimum value as in the case with AGN (the minimum value of $r_{\rm{cool,NoAGN}}/R_{200}$ is 0.4 compared with 0.08 for $r_{\rm{cool,AGN}}/R_{200}$). This means that the rate of gas cooling from the hot halo is always larger in the absence of AGN activity.   

We find that $\dot{M}_{\rm{gas,galaxy}}$(no AGN) steadily increases because the rate of gas cooling from the hot halo, $\dot{M}_{\rm{cooling}}/\dot{M}_{\rm{gas,halo}}=(r_{\rm{cool}}/R_{200})f^{\rm{halo}}_{\rm{acc,hot}}$, remains roughly constant with increasing halo mass. We obtain that $(r_{\rm{cool}}/R_{200})f^{\rm{halo}}_{\rm{acc,hot}}\sim 0.4$ for haloes with masses between $10^{12}-10^{13}\Msun$ ($f^{\rm{halo}}_{\rm{acc,hot}}$ increases with halo mass at roughly the same rate as $r_{\rm{cool}}/R_{200}$ decreases). However, since $f^{\rm{halo}}_{\rm{acc,cold}}$ decreases with halo mass, $\dot{M}_{\rm{gas,galaxy}}$(no AGN) differs from $\epsilon\dot{M}_{\rm{gas,halo}}$ by 0.1 dex for $10^{12}\Msun$ haloes and 0.3 dex for $10^{13}\Msun$ haloes. Physically, the steady increase of $\dot{M}_{\rm{gas,galaxy}}$(no AGN) with halo mass means that when there is no AGN feedback, a larger rate of gas cooling from the hot halo develops. We conclude that the formation of a hot halo alone is not enough to prevent gas from reaching the galaxy (see also \citealt{Gabor11, Liu17, Gutke17}).

%Hot halo quenching alone has been shown to be unable to fully suppress star formation (e.g. \citealt{Liu17}) or reproduce observed luminosity functions of red galaxies without additional external heating (e.g. \citealt{Gabor11}). It can only quench a galaxy without the need for AGN feedback if gas is prevented from cooling during the last 10Gyr (\citealt{Gutke17}).

\section{Summary and conclusions}\label{Conclusion_sec}

We have investigated the physics that drives the gas accretion rates onto galaxies at the centers of dark matter haloes using the EAGLE simulation suite as well as analytic calculations. We began by defining the gas accretion rate onto the galaxy as the rate at which gas crosses the radius $0.15\times R_{200}$ between two consecutive snapshots. We also defined the gas accretion rate onto the ISM as the rate 	at which gas crosses the radius $0.15\times R_{200}$ and the phase space cut $n_{\rm{H}}>0.1\rm{cm}^{-3}$, $T<10^{5}\rm{K}$. We found that at $z=0$ and in the halo mass range $10^{10}-10^{12}\Msun$ the gas accretion rates onto the galaxy ($\dot{M}_{\rm{gas,galaxy}}$) and ISM ($\dot{M}_{\rm{ISM}}$) increase with halo mass at approximately the same rate, with $\dot{M}_{\rm{ISM}}$ having a 0.3 dex (on average) lower normalization than $\dot{M}_{\rm{gas,galaxy}}$. For halo masses $\gtrsim 10^{12}\Msun$, $\dot{M}_{\rm{ISM}}$ remains nearly constant. While $\dot{M}_{\rm{gas,galaxy}}$ flattens at ${\sim}10^{12}\Msun$, it increases with halo mass for haloes with masses $\gtrsim 10^{13}\Msun$ (Fig. 1).

We analysed the dependence of the rates of gas accretion onto galaxies, as well as onto haloes, on halo mass and redshift. We defined the rate of gas accretion onto haloes ($\dot{M}_{\rm{gas,halo}}$) as the rate at which gas crosses the virial radius between two consecutive snapshots, and compared the simulation output with the analytic prediction of \citet{Correa15c} for the dark matter accretion rate, scaled by the universal baryon fraction. \citet{Correa15b,Correa15c} demonstrated that their analytic model reproduces the accretion rates onto haloes in collisionless simulations. We found that the analytic prediction is 0.8 dex too high for $10^{10}\Msun$ haloes at $z=0$, agrees for $10^{11.5}\Msun$ haloes, but is 0.3 dex too low for $10^{13}\Msun$ haloes (Fig. 2). At redshifts $z=1-2$ better agreement is obtained between $\dot{M}_{\rm{gas,halo}}$ and the analytic prediction for haloes with mass $\gtrsim 10^{12}\Msun$. At higher redshifts ($z>2$) the analytic prediction and $\dot{M}_{\rm{gas,halo}}$ increase with halo mass at the same rate. However, the analytic prediction has a higher normalization (by up to $\sim 0.5$ dex for $z\sim 8$). We believe that better agreement would be reached for high-mass haloes if the analytic model included the impact of baryonic physics that reduces the halo mass. The large discrepancy for halo masses of $\sim 10^{10}\Msun$ is expected because the potential wells of these haloes are too shallow to strongly bind photo-heated gas (e.g. \citealt{Sawala13,Benitez17}).

The gas accretion rate onto galaxies is (on average) a factor of 4 (and up to a factor of 16) lower than that onto haloes for halo masses $10^{10}-10^{12}\Msun$ ($\sim 10^{12.5}\Msun$) at $z=0$, and a factor of 2 (5) lower for $10^{10}-10^{12}\Msun$ ($\sim 10^{12.5}\Msun$) haloes at $z=2$. In low-mass haloes, $\dot{M}_{\rm{gas,galaxy}}$ increases with halo mass at nearly the same rate as $\dot{M}_{\rm{gas,halo}}$, but it flattens for halo masses of $\sim 10^{12}\Msun$ at $z=0-2$ (Fig. 2). In high-mass haloes ($>10^{12}\Msun$), two modes of accretion (hot and cold) coexist and contribute to the total rates onto galaxies and haloes. We defined hot gas accretion as the accretion rate of gas that after accretion onto the galaxy or halo has a temperature higher than $10^{5.5}\rm{K}$, and calculated the fraction of gas accreted hot onto the galaxy ($f^{\rm{galaxy}}_{\rm{acc,hot}}$) and halo ($f^{\rm{halo}}_{\rm{acc,hot}}$). We found that while $f^{\rm{halo}}_{\rm{acc,hot}}=0.7$ for $10^{12}\Msun$ haloes at $z=0$, $f^{\rm{galaxy}}_{\rm{acc,hot}}=0.05$ for the central galaxies within them. However, this difference decreases for higher-mass haloes, e.g. $f^{\rm{halo}}_{\rm{acc,hot}}=0.98$ and $f^{\rm{galaxy}}_{\rm{acc,hot}}=0.80$ for $10^{13}\Msun$ haloes at $z=0$. The fraction of gas accreted hot onto galaxies strongly depends not only on halo mass but also on redshift. We found that $\dot{M}_{\rm{gas,galaxy}}$ changes from being cold-mode dominated to hot-mode dominated ($f^{\rm{galaxy}}_{\rm{acc,hot}}=0.5$) for galaxies in $10^{12.7}\Msun$ haloes at $z=0$ but in $10^{13.3}\Msun$ haloes at $z=1$ (Fig. 3).

We also investigated the dependence of the rates of gas accretion onto galaxies on feedback variations. It has been shown that in scenarios with energetic stellar feedback, the galaxy accretion rates can increase due to re-accretion of gas that was blown out of the galaxy by stellar-driven winds. In this work, we found that when stellar feedback is twice as energetic, $\dot{M}_{\rm{gas,galaxy}}$ for galaxies in $\leq 10^{12}\Msun$ haloes is lower than $\dot{M}_{\rm{gas,galaxy}}$ from Ref. However we find that $\dot{M}_{\rm{gas,galaxy}}$ increases for galaxies in haloes larger than $10^{12}\Msun$, indicating that at these halo masses galaxies re-accrete gas that was ejected by stellar feedback in lower-mass progenitors. When stellar feedback is half as energetic, $\dot{M}_{\rm{gas,galaxy}}$ decreases for galaxies in $\geq 10^{11}\Msun$ haloes, possibly due to a lower rate of re-accreted gas or because black hole accretion, and hence AGN feedback, becomes efficient at lower halo masses (Fig 4).

When stellar feedback is turned off, $\dot{M}_{\rm{gas,galaxy}}$ decreases relative to Ref for galaxies in $>10^{10.5}\Msun$ haloes. This is likely because without stellar feedback the central black holes in low-mass galaxies become much more massive (\citealt{Bower17}), allowing AGN feedback to suppress the gas infall rates. Indeed, when both stellar and AGN feedback are turned off, the rates of gas accretion are higher than in Ref for galaxies in $\lesssim 10^{12}\Msun$ haloes (Fig. 4).

When AGN feedback is more explosive and intermittent, $\dot{M}_{\rm{gas,galaxy}}$ decreases with respect to Ref for galaxies in $\geq 10^{12}\Msun$ haloes, indicating that AGN further suppress $\dot{M}_{\rm{gas,galaxy}}$ in massive haloes. When AGN feedback is switched off, $\dot{M}_{\rm{gas,galaxy}}$ does not flatten at $\sim 10^{12}\Msun$ and increases with halo mass at the same rate as the gas accretion onto haloes but with a lower normalization (Fig. 5). This shows that AGN feedback is the mechanism responsible for preventing hot gas from cooling and falling onto central galaxies in massive haloes.

To further understand the behavior of $\dot{M}_{\rm{gas,galaxy}}$ with halo mass, we developed a physically motivated semi-analytic model of galaxy gas accretion. The model postulates that two modes of accretion, cold and hot, contribute to the total gas accretion rate. Cold gas accretion onto galaxies is driven by the rate of cold accretion onto haloes, whereas hot gas accretion is driven by the rate of gas cooling from the hot halo, which depends on the rate of gas accreting onto the halo that is shock-heated and on the location of the cooling radius.

To calculate the cooling radius, $r_{\rm{cool}}$, we equated the heating rate produced by accretion shocks (derived in Paper I) with the cooling rate. We found that in the radial range $[0.1-1]R_{200}$ the hot gas density profile of haloes from the EAGLE simulations deviates from the isothermal shape. The logarithmic density slope increases with halo mass for $>10^{12}\Msun$ haloes, reaches a maximum of $-0.9$ in $10^{13}\Msun$ haloes, and decreases at higher masses. The change in the slope of the density profile reflects how the distribution of the hot gas changes as the halo evolves due to continued infall, reheating and cooling. Because the density profile evolves with halo mass, the ratio $r_{\rm{cool}}/R_{200}$ does not decrease monotonically with halo mass. It decreases up to ${\sim}10^{13}\Msun$ haloes and increases towards larger haloes (Fig. 6). 

The upturn in the ratio of the cooling and virial radii with halo mass may indicate that while AGN-driven outflows reduce the density of the hot halo for $\sim 10^{12}-10^{13}\Msun$ haloes, in higher-mass haloes AGN feedback is insufficiently efficient, causing the hot halo cooling flow to become more prominent, and the accretion rate onto galaxies to increase more steeply with halo mass. When there is no AGN feedback, the density of the hot halo is higher, $r_{\rm{cool}}/R_{200}$ does not decrease as much as when AGN feedback is on, and so the rate of gas cooling from the hot halo is higher. We compared our semi-analytic model of gas accretion with the galaxy accretion rates calculated from the simulation at $z=0$ and found excellent agreement.

%AGN-driven outflows reduce the density of the hot halo for $\sim 10^{12}-10^{13}\Msun$ haloes, thus decreasing $r_{\rm{cool}}/R_{200}$ and flattening the dependence of the accretion rate onto galaxies on halo mass. In higher-mass haloes, the cooling radius increases, the hot halo cooling flow becomes more prominent, and the accretion rate onto galaxies increases more steeply with halo mass. This may however be a result of insufficiently efficient AGN feedback. When there is no AGN feedback, the density of the hot halo is higher, $r_{\rm{cool}}/R_{200}$ does not decrease as much as when AGN feedback is on, and so the rate of gas cooling from the hot halo is higher. We compared our semi-analytic model of gas accretion with the galaxy accretion rates calculated from the simulation at $z=0$ and found excellent agreement. 

In future work, we plan to investigate whether the correlation between accretion rates onto nearby galaxies (i.e. galactic conformity, \citealt{Weinmann06}) is driven by the correlation of dark matter halo formation time with environment. We will test the predictions of the semi-analytic model with the simulation outputs.

%The semi-analytic model derived in this work relates $\dot{M}_{\rm{gas,galaxy}}$, as well as the hot halo cooling flow (through the calculation of $r_{\rm{cool}}$), to $\dot{M}_{\rm{gas,halo}}$. Interestingly, if we consider two halos (A and B) of the same mass but with different formation times (with halo A forming earlier than halo B, likely because of residing in a denser environment), the haloes will contain galaxies of different stellar masses (due to galactic downsizing, e.g. Neistein et al. 2006) and will have different rates of gas accretion (by up to a factor of 2 for $10^{12}\Msun$ halos if the formation time is varied by a factor of 1.5, according to the semi-analytic model of Correa et al. 2015b). When we include these differences in the semi-analytic model for $\dot{M}_{\rm{gas,galaxy}}$, we obtain that the accretion rates onto their central galaxies are nearly the same, suggesting that the correlation between accretion rates onto nearby galaxies (i.e. galactic conformity, \citealt{Weinmann06}) is not necessarily driven by the correlation of dark matter halo formation time with environment. We aim to explore the validity of this prediction of our semi-analytic model in future work.

%Recently, \citet{vandeVoort17} calculated the rates of gas accretion onto central galaxies as a function of stellar mass for fixed halo mass (Fig. 2 of van de Voort 2017), and obtained constant rates of accretion onto galaxies as a function of stellar mass for halos more massive than $10^{12}\Msun$.

\section*{Acknowledgments}

We are grateful to the EAGLE team for putting together a great set of simulations. This work used the DiRAC Data Centric system at Durham University, operated by the Institute for Computational Cosmology on behalf of the STFC DiRAC HPC Facility (www.dirac.ac.uk). This equipment was funded by BIS National E-infrastructure capital grant ST/K00042X/1, STFC capital grant ST/H008519/1, and STFC DiRAC Operations grant ST/K003267/1 and Durham University. DiRAC is part of the National E-Infrastructure. The EAGLE simulations were performed using the DiRAC-2 facility at Durham, managed by the ICC, and the PRACE facility Curie based in France at TGCC, CEA, Bruyeres-le-Chatel. This work was supported by the Netherlands Organisation for Scientific Research (NWO) through VICI grant 639.043.409. FvdV acknowledges the Klaus Tschira Foundation. We thank the anonymous reviewer for fruitful comments that improved the original manuscript.

\bibliography{biblio}
\bibliographystyle{mn2e}

\appendix

%\begin{table*}
%\centering  %r@{.}l
%\caption{List of simulations.  From
%  left-to-right the columns show: simulation identifier; comoving box size;
%  number of dark matter particles (there are equally 
%  many baryonic particles); initial baryonic particle mass; dark matter
%  particle mass; comoving (Plummer-equivalent) gravitational
%  softening and maximum physical softening.} 
%\label{Table_sims}
%\begin{tabular}{lrrccrrc}
%\hline
%  Simulation & L & N & $m_{\rm{b}}$ & $m_{\rm{dm}}$ & $\epsilon_{\rm{com}}$ & $\epsilon_{\rm{prop}}$ \\ 
%   & ($\rm{cMpc}$) & & ($\rm{M}_{\sun}$) & ($\rm{M}_{\sun}$) & ($\rm{ckpc}$) & ($\rm{pkpc}$) \\  \hline\hline
%   L100N1504 & 100 & $1504^{3}$ & $1.81\times 10^{6}$ & $9.70\times 10^{6}$ & 2.66 & 0.70 \\ 
%   L100N0752 & 100 & $752^{3}$ & $1.44\times 10^{7}$ & $7.76\times 10^{7}$ & 5.32 & 1.40 \\ 
%   L050N0752 & 50 & $752^{3}$ & $1.81\times 10^{6}$ & $9.70\times 10^{6}$ & 2.66 & 0.70 \\ 
%   L050N0376 & 50 & $376^{3}$ & $1.44\times 10^{7}$ & $7.76\times 10^{7}$ & 5.32 & 1.40 \\ 
%   L025N0752 & 25 & $752^{3}$ & $2.26\times 10^{5}$ & $1.21\times 10^{6}$ & 1.33 & 0.35 \\
%   L025N0376 & 25 & $376^{3}$ & $1.81\times 10^{6}$ & $9.70\times 10^{6}$ & 2.66 & 0.70\\ \hline
%\end{tabular}
%\end{table*}

\section{Simulations}\label{Simulations_app}

In the EAGLE simulations, star formation is modeled following the recipe of \citet{Schaye08}. It is stochastic above a density threshold that depends on metallicity (proposed by \citealt{Schaye04}). Stellar evolution and mass loss follows the work of \citet{Wiersma09b}, where star particles are treated as simple stellar populations with \citet{Chabrier} initial mass function, spanning the range $0.1-100 \Msun$. Feedback from star formation follows the stochastic thermal feedback scheme of \citet{DallaVecchia12}. Rather than heating all neighbouring gas particles within the SPH kernel, they are selected stochastically based on the available energy, then heated by a fixed temperature difference of $\Delta T = 10^{7.5}$K. The probability that a neighbouring SPH particle is heated is determined by the fraction of the energy budget that is available for feedback, that depends on adjustable maximum and minimum threshold values ($f_{\rm{th,max}}$ and $f_{\rm{th,min}}$, respectively), on the gas density as well as on metallicity.

For AGN feedback, black hole seeds (of $\approx 1.4\times 10^{5}\Msun$) are included in haloes with mass greater than $\approx 1.4\times 10^{10}\Msun$ (\citealt{Springel05}). Black holes can grow through mergers and accretion. The accretion events follow a modified Bondi-Hoyle formula that accounts for the angular momentum of the accreting gas (\citealt{Rosas13}), and a free parameter that is related to disk viscosity ($C_{\rm{visc}}$). AGN feedback is stochastic, it follows the accretion of mass onto the black hole, where a fraction of the accreted gas is released as thermal energy into the surrounding gas. This method is based on that of \citet{Booth} and \citet{DallaVecchia12}, where the free parameter is the heating temperature $\Delta T_{\rm{AGN}}$. Finally, radiative cooling and photo-heating are included as in \citet{Wiersma09b}. The element-by-element radiative rates are computed in the presence of the cosmic microwave background (CMB) and the \citet{Haardt} model for UV and X-ray background radiation from quasars and galaxies. 

The EAGLE simulations include a new SPH formulation named `Anarchy' which improves the performance on standard hydrodynamical tests when compared to the original SPH implementation in GADGET (see \citealt{Schaller15b} or \citealt{Hu14} for similar results). Anarchy makes use of a pressure-entropy formulation derived in Hopkins (2013), allowing it to avoid spurious jumps at contact discontinuities. It also uses an artificial viscosity switch as in \citet{Cullen10}, that allows the viscosity limiter to be stronger when shocks and shear flows are present. In addition, it includes an artificial conduction switch (similar to that of \citealt{Price08}), the $C^{2}$ \citet{Wendland95} kernel and the time step limiters of \citet{Durier12}, which ensure that ambient particles do not remain inactive when a shock is approaching (for a more complete description see \citealt{Schaye14}).

\section{Numerical convergence}\label{Numerical_convergence}

\begin{figure} 
  \centering
  \includegraphics[angle=0,width=0.46\textwidth]{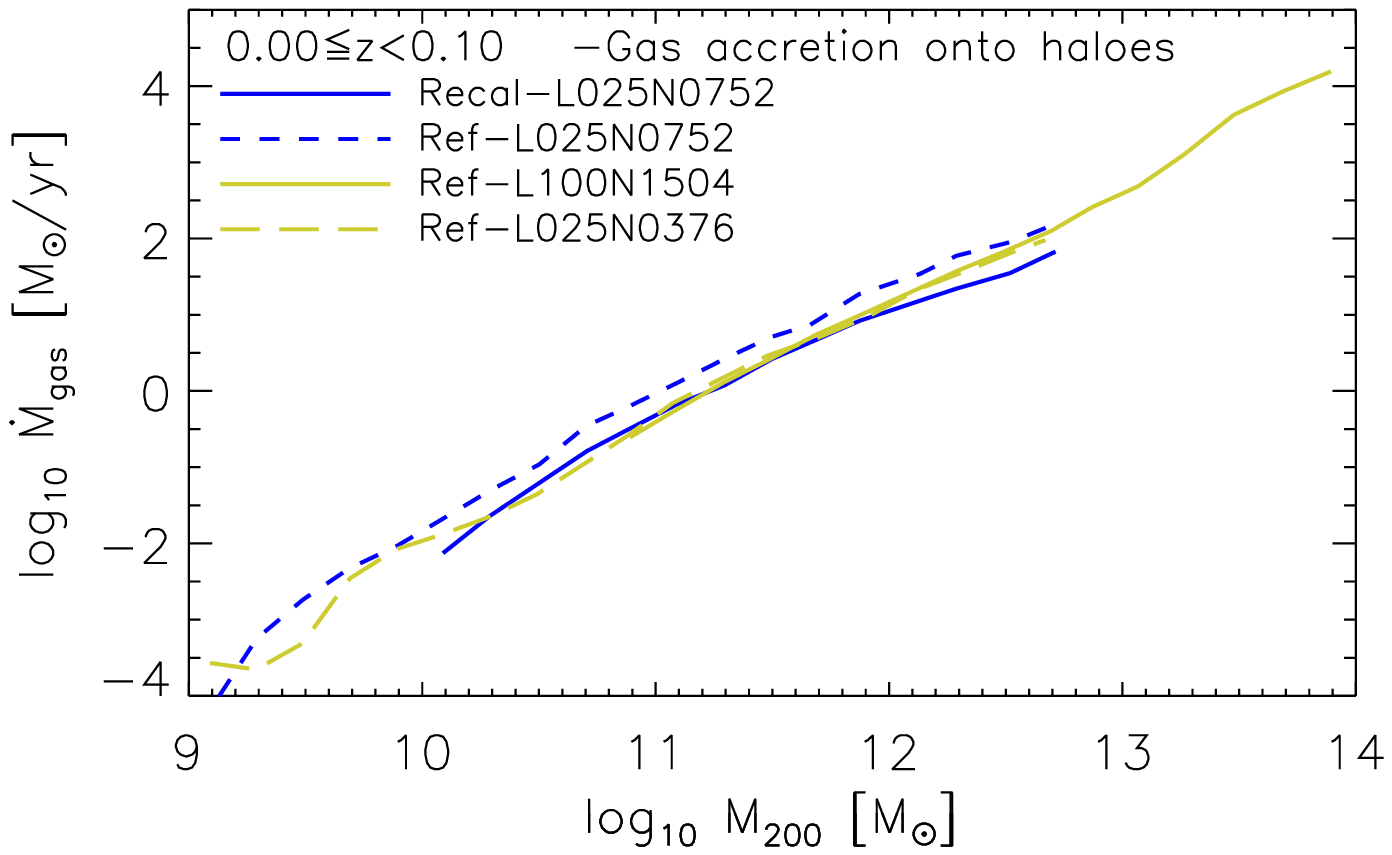}\\
  \vspace{-0.2cm}
  \includegraphics[angle=0,width=0.46\textwidth]{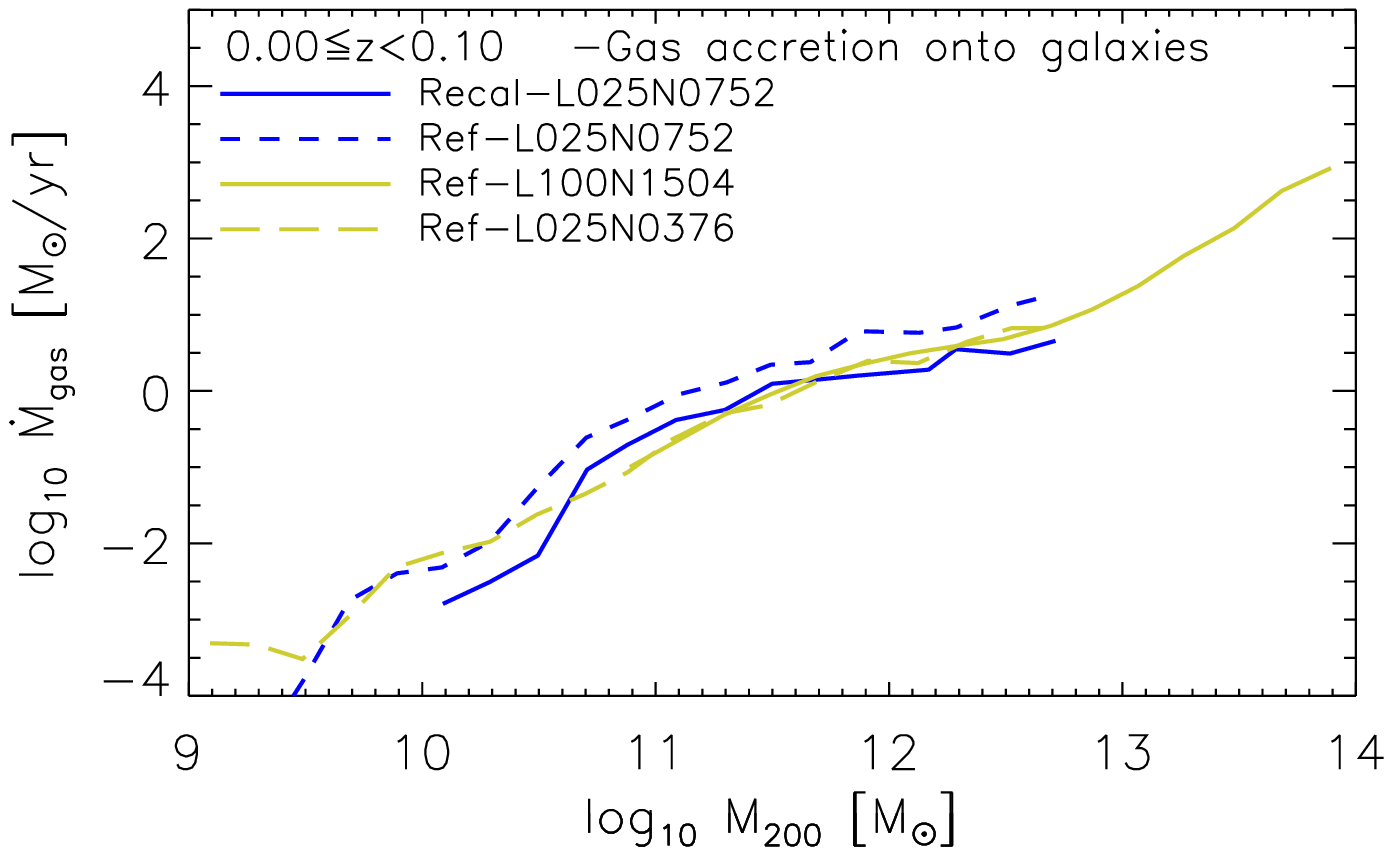}
  \caption{Accretion rate of gas onto haloes (top panel) and their central galaxies (bottom panel) as a function of halo mass in the redshift range $0\leq z<0.1$. The curves show the median values of the total accretion rates of haloes, as well as onto galaxies, in logarithmic mass bins of 0.2 dex, each mass bin contains at least 10 haloes. To analyze numerical convergence, we compare accretion rates from simulations with different box sizes and number of particles and therefore different resolution. We classify the simulations as high-resolution (blue curves) and intermediate-resolution (yellow curves). We find strong convergence with box size and weak convergence between different resolution simulations (see text for details).}
\label{accretion_convergence}
\end{figure}

%We find excellent agreement between the accretion rates onto haloes and galaxies from simulations with same resolution and different box size, but we do not achieve `strong convergence', which means agreement between simulations with different numerical resolution and same parameters of the subgrid models. In the latter case we find that in $10^{11}\Msun$ haloes, $\dot{M}_{\rm{gas,galaxy}}$ increases by up to a factor of 10 if the mass resolution is increased by a factor of 64, whereas $\dot{M}_{\rm{gas,halo}}$ increases by up to a factor of 2.5. This is not discouraging because we achieve `weak convergence', meaning that we find excellent agreement in the accretion rates between the recalibrated model `Recal-L025N0752' and the reference models Ref-L100N1504/Ref-L025N0376. As explained in \citet{Schaye14}, although strong convergence is a necessary condition for predictive power, only weak convergence is required for observables that depend strongly and directly on the efficiency of subgrid feedback. For a detailed analysis of numerical convergence see Appendix~\ref{Numerical_convergence}. 

We investigate how numerical resolution impacts the rates of gas accretion in the EAGLE simulations. Fig.~\ref{accretion_convergence} shows the total gas accretion onto haloes (top panel) and their central galaxies (bottom panel) in the redshift range $0\leq z<0.1$. Both panels show that the accretion rates (onto haloes and galaxies) increase with increasing halo mass. We find that in $10^{11}\Msun$ haloes, $\dot{M}_{\rm{gas,galaxy}}$ increases (on average) by a factor of 3 if the  particle mass resolution is increased by a factor of 8, whereas $\dot{M}_{\rm{gas,halo}}$ increases by up to a factor of 2.5. This lack of convergence in the gas accretion rates onto galaxies is expected, due to the fact that recycling winds from SN and AGN outflows are resolution dependent. 

%For simulations that do no resolve the ISM, numerical resolution is directly related to the efficiency of feedback mechanisms which gives rise to the need for (re-)calibration (see \citealt{Schaye14} for a discussion). 
However, we achieve convergence between simulations that use different resolution but for which the parameters of the subgrid feedback have been calibrated to the same observations. We find excellent agreement in the accretion rates between the recalibrated model `Recal-L025N0752' and the reference models Ref-L100N1504/Ref-L025N0376. %As explained in \citet{Schaye14}, although strong convergence is a necessary condition for predictive power, only weak convergence is required for observables that depend strongly and directly on the efficiency of subgrid feedback. 

The top panel from Fig.~\ref{accretion_convergence} shows a drop in the accretion rates from the intermediate-resolution simulations (yellow curves) in haloes smaller than $10^{10}\Msun$. This is likely a numerical artifact because such a drop is not seen at this mass in the accretion rates from the high resolution simulation (blue curve). We therefore set the minimum halo mass for accretion onto haloes (and their inner galaxies) to correspond to 1000 dark matter particles ($\sim 10^{10}\Msun$ halo mass for the intermediate-resolution simulations). Throughout this work we use the intermediate-resolution (fiducial) simulations.

%\subsection{Impact of hydrodynamic scheme}

%Finally, we compare simulations with different hydrodynamics schemes. In Paper I we discuss the differences between the standard SPH code GADGET and a more recent formulation of SPH included in the EALGE simulations named Anarchy (Dalla Vecchia 2016 in prep.). We find that while gas mixing is largely suppressed in GADGET, Anarchy is able to mix phases in contact discontinuity allowing dense clumps to dissolve into the hot halo. As a result the fraction of hot gas accretion onto haloes is lower (by up to 0.2 dex) in the GADGET simulation with respect to its Anarchy counterpart. In the bottom panel of Fig.~\ref{gal_accr_comparison} we compare the gas accretion rates onto galaxies calculated from the Ref (orange solid line), the NoFeedback/Anarchy (purple short-dashed line) and NoFeedback/GADGET (blue long-dashed line) simulations. The panel shows that in the absence of feedback the gas accretion rates onto galaxies is larger (by up to an order of magnitude) in haloes smaller than $10^{11.5}\Msun$. In larger haloes we find very good agreement between accretion rates. In this last case we caution the reader that we are comparing same-resolution simulations run in the L025N0376 box, as a result the number of haloes above $10^{12}\Msun$ is $\sim$10 in each mass bin and therefore it does not represent a complete sample.

\section{Density profiles of hot gaseous haloes}\label{density_profile}

In this section we investigate how the density profile of hot halo gas evolves due to the continued infall, reheating and cooling of gas in haloes from the EAGLE simulations.

We select all haloes from the Ref-L100N1504 and No AGN FB-L050N0752 simulations and separate them in mass bins of $\Delta \log_{10}(M_{200})=0.2$ width. To calculate $\rho_{\rm{hot-gas}}(r)$, we select gas particles that are hot (have cooling times longer than local dynamical times) and define a set of concentric spherical shells of width $\Delta\log_{10}(r)=0.078$. We add the mass of the particles within each shell and divide by the volume. As we are only interested in the density profile of gas in the haloes but not within galaxies, we restrict the analysis of $\rho_{\rm{gas}}(r)$ to the radial range $0.1$ to $1\times R_{200}$.

We calculate the dynamical time, $t_{\rm{dyn}}$, of the gas particle as $t_{\rm{dyn}}=r/V_{\rm{c}}(r)$, where $V_{\rm{c}}(r)=[GM(<r)/r]^{1/2}$ is the circular velocity and $M(<r)$ is the mass enclosed within $r$. We calculate the cooling time, $t_{\rm{cool}}$, as $t_{\rm{cool}} = \frac{3nk_{\rm{B}}T}{2\Lambda}$, where $n$ is the number density of the gas particle ($n=\rho_{\rm{gas}}/\mu m_{\rm{p}}$, $\mu m_{
rm{p}}$ is the mean particle mass calculated from the cooling tables of \citet{Wiersma09b}), $k_{\rm{B}}$ is the Boltzmann constant, $T$ is the gas temperature and $\Lambda$ is the cooling rate per unit volume with units of erg cm$^{-3}$s$^{-1}$. To calculate $\Lambda$, we use the tabulated cooling function for gas exposed to the evolving UV/X-ray background from \citet{Haardt} given by \citet{Wiersma09b}, which was also used by the EAGLE simulations. Note that the ``standard" definition for the dynamical time of gas within a virialized system depends on $R_{200}$ and $V_{\rm{c}}(R_{200})$, and not on the local radius and local circular velocity as defined here. 

\begin{figure} 
 \centering
  \subfloat{\includegraphics[angle=0,width=0.45\textwidth]{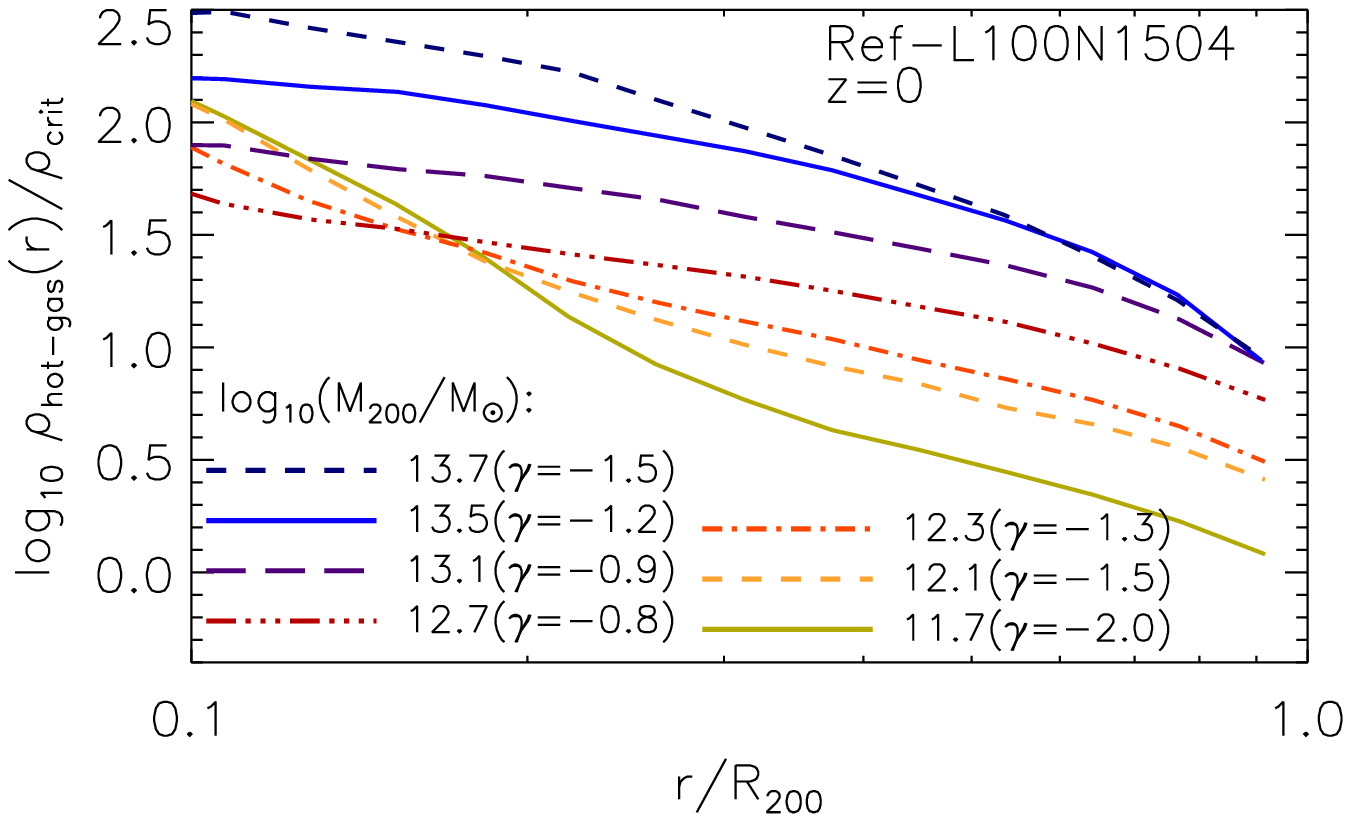}}\\
  \vspace{-0.7cm}
  \subfloat{\includegraphics[angle=0,width=0.45\textwidth]{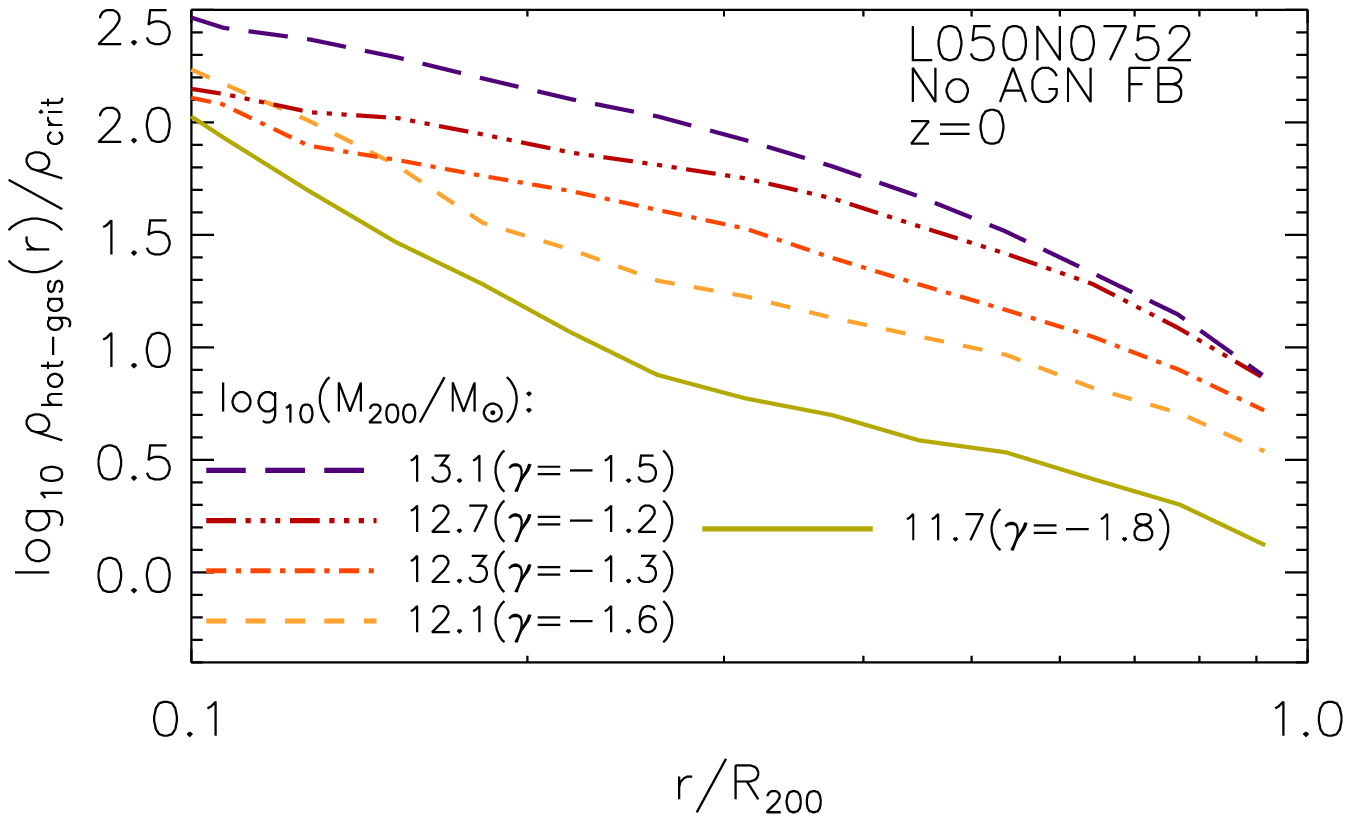}}
  \vspace{0.1cm}
  \caption{Average density profile of hot gas at $z=0$ of stack of haloes in the mass range $\log_{10}(M_{200}/\Msun)\pm 0.1$, with $\log_{10}(M_{200}/\Msun)$ varying from 11.7 to 13.7 (as indicated in the legends) from the Ref model (top panel) and the No AGN FB model (bottom panel).}
\label{density_plot_1}
\end{figure}

Fig.~\ref{density_plot_1} shows the median gas density profile at $z=0$ of haloes in the mass range $\log_{10}(M_{200}/\Msun)\pm 0.1$, with $\log_{10}(M_{200}/\Msun)$ varying from 11.7 to 13.7, as indicated in the legends, from the Ref model (top panel) and from the No AGN FB model (bottom panel). We define $\gamma$ as the logarithmic density slope (${\rm{d}}\ln \rho_{\rm{gas}}/{\rm{d}}\ln r\equiv\gamma$), which we measure between the radial range 0.1 and $1\times R_{200}$, assuming $\rho_{\rm{hot-gas}}(r)\propto r^{\gamma}$. We find that $\gamma$ increases from $-1.8$ in $10^{12}\Msun$ haloes to $-0.8$ in $10^{12.7}\Msun$ haloes, and then decreases again towards larger halo masses. In smaller haloes, $\rho_{\rm{hot-gas}}(r)$ is steeper at small radii, and deviates from the isothermal shape (characterized by $\gamma=-2$).

The change in the slope of $\rho_{\rm{hot-gas}}(r)$ with mass indicates how the distribution of the hot gas in the halo evolves, and can be explained as follows. Since a stable hot halo forms for $\sim 10^{12}\Msun$, most gas crossing $R_{200}$ remains hot. As a result the gas density in the range $0.5-1R_{200}$ increases and the density profile flattens. When AGN feedback is turned off,  $\rho_{\rm{hot-gas}}(r)$ does not flatten as much as in Ref, suggesting that the cooling flow from the hot halo is more prominent.

\section{Accretion rate onto haloes}\label{Comparison}

\begin{figure} 
 \centering
  \subfloat{\includegraphics[angle=0,width=0.4\textwidth]{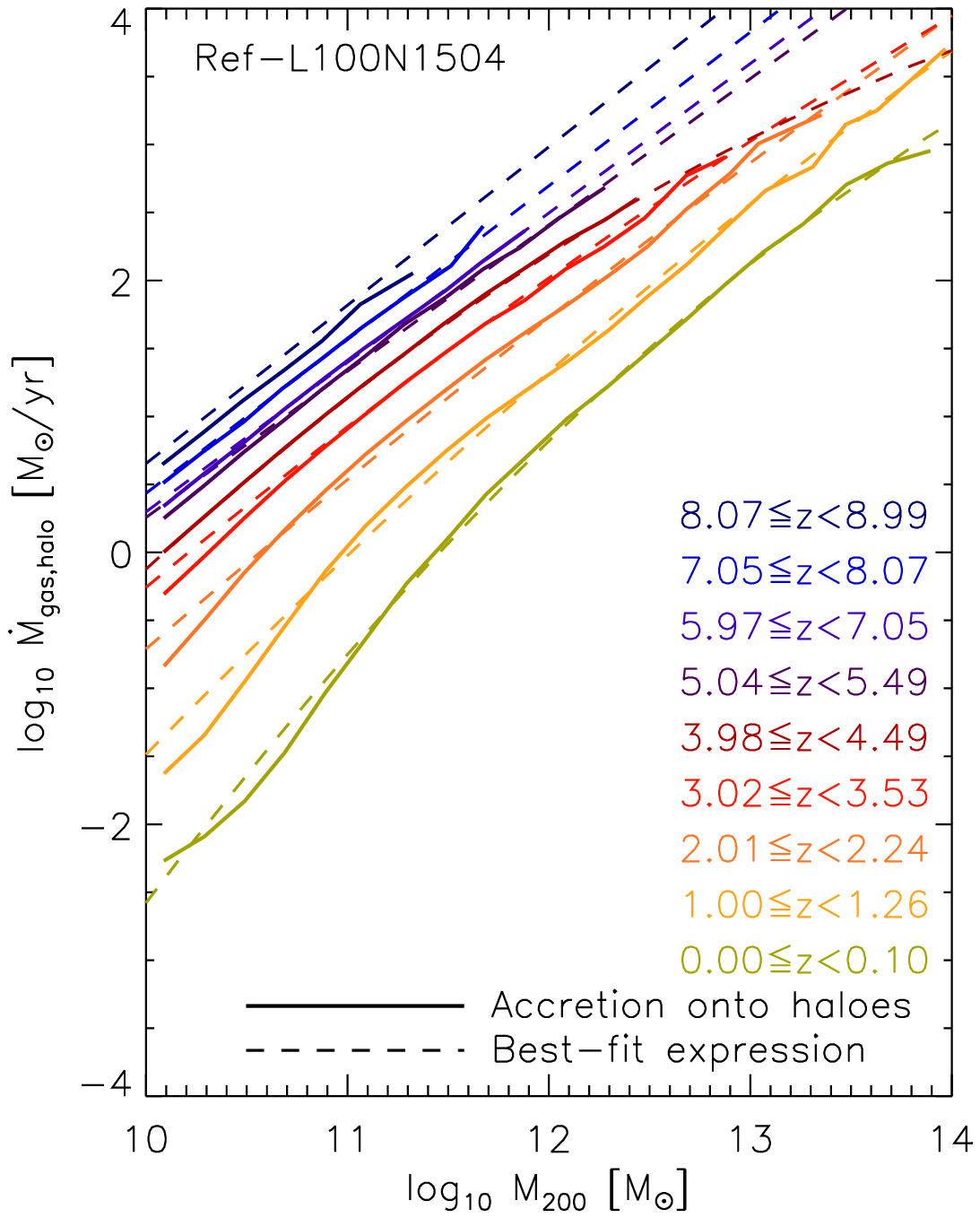}}
  \caption{Accretion rates onto haloes as a function of halo mass for different redshifts. The curves are colored according to the redshift intervals indicated in the legends. The solid curves correspond to the accretion rates taken from the Ref-L100N1504 simulation, whereas the dashed curves correspond to the best-fit expression presented in Section \ref{accretion_rates_halos_galaxies} given by eqs. (1-8).}
\label{comparison_plot}
\end{figure}

In this section we show a comparison between the best-fit relations presented in Section~\ref{accretion_rates_halos_galaxies} given by eqs. (1-8) and the simulation outputs. Fig.~\ref{comparison_plot} shows the accretion rates onto haloes as a function of halo mass for different redshifts. The solid curves correspond to the accretion rates taken from the Ref-L100N1504 simulation, whereas the dashed curves correspond to the best-fit expression. We show that the best-fit expressions are able to closely reproduce the gas accretion rate onto haloes in the redshift range 0 to 8 and halo mass range $10^{10}$ to $10^{14}\Msun$.

\end{document}